\documentclass{aa}
\usepackage{graphicx}
\usepackage{txfonts}
\usepackage{natbib}
\usepackage{lscape}
\usepackage{arydshln}

\bibpunct{(}{)}{;}{a}{}{,}
\usepackage{float}
\usepackage{appendix}
\usepackage{amsmath}
\usepackage[switch]{lineno}

\begin{document} 
% \linenumbers

   \title{Multiple stellar populations at less evolved stages. IV. evidence of helium enrichments in four Magellanic globular clusters} 

   \subtitle{}

   \author{Chengyuan Li\inst{1,2}\fnmsep\thanks{Corresponding author.}, 
    Li Wang\inst{1,2}, Xin Ji\inst{3}, Holger Baumgardt\inst{4}}
    \institute{School of Physics and Astronomy, Sun Yat-sen University, Zhuhai 519082, China \\
    \email{lichengy5@mail.sysu.edu.cn} \and
    CSST Science Center for the Guangdong-Hongkong-Macau Greater Bay Area, Sun Yat-sen University, Zhuhai 519082, China \and
    National Astronomical Observatories, Chinese Academy of Sciences, Beijing 100012, China \and School of Mathematics and Physics, The University of Queensland, St. Lucia, QLD 4072, Australia}

   \date{Submitted: \today }

% \abstract{}{}{}{}{} 
% 5 {} token are mandatory
 
  \abstract
  % context heading (optional)
  % {} leave it empty if necessary  
   {Most globular clusters in the local group show multiple stellar populations, a pattern defined by variations of several light elements, with some also displaying dispersions in heavy elements. Since helium is the most immediate product of hydrogen burning, almost all models suggest that second-generation stars should show different levels of helium enrichment than first-generation stars. Therefore, investigating helium enrichment in stellar populations of globular clusters can constrain different theoretical models. Using the deep photometry carried out by the {\sl Hubble} Space Telescope, we have analyzed the morphologies of the main sequences of four Large Magellanic Cloud globular clusters, Hodge 11, NGC 1841, NGC 2210, and NGC 2257. We aim to constrain the helium distributions among their main sequence stars. We found a clear broadening of main sequences in all four clusters. After excluding the effects of photometric errors, differential reddening, unresolved binaries, and metallicity dispersions, this broadening would suggest a substantial helium enrichment in their star populations, ranging from $\delta{Y}=0.08$ to $\delta{Y}\geq0.12$, depending on the presumed helium distributions. Helium-enriched stars are comparable in number to normal stars within these clusters, and may even dominate, comprising approximately $\sim$40\% to over 80\% of the total stellar population. However, given the great distance of the Large Magellanic Cloud, it is difficult to rule out the significant impact of crowding, so our results may only represent an upper limit.}
   
   % conclusions heading (optional), leave it empty if necessary 
   
   \titlerunning{LMC GCs}
   \authorrunning{Li et al.}

   \maketitle
%

%%==============================================================================================================================%%
\section{Introduction}\label{S1}
Owing to photometric and spectroscopic observations conducted with various ground-based and space telescopes, recent studies have revealed that nearly all globular clusters (GCs) within the Local Group exhibit multiple stellar populations (MPs). The phenomenon of MPs is characterized by the fact that nearly all member stars within GCs exhibit abundance dispersions, involving light elements such as He, C, N, O, Na, Mg, and Al, and some even exhibit dispersions in heavy elements \citep[see][for an early and a more recent reivews]{Grat04a,Milo22a}. As expected, the phenomenon of MPs was first, and most extensively, reported in the GCs of the Milky Way \citep[e.g.,][for an early review]{Smit87a}. Subsequently, since last decade, it was discovered that star clusters with MPs also exist in the satellite galaxies, indicating that the MPs may be widespread among star clusters in different galaxies \citep[][for examples in Fornax dSph and Magellanic Clouds]{Lars12a,Milo20a}.

The nature of the polluters remains debated, including asymptotic giant branch (AGB) stars \citep[e.g.,][]{Derc08a,Sies10a,Dohe14a}, fast-rotating massive stars \citep[FRMS,][]{Decr07a,Krau13a}, very massive stars \citep[VMS,][]{Vink18a,Giel18a}, and interactive binaries \citep[IBs,][]{Demi09a,Jian14a,Nguy24a}, or some of their combinations \citep[e.g.,][]{Wang20a,Wint23a}. For any scenario aimed at explaining elemental enrichment, given that helium is the most direct product of stellar nucleosynthesis, the polluters proposed by these models must also exhibit He-enrichment. Most importantly, these models would indicate different relationships between the dispersions of light elements and He abundance. Thus, studying He abundance distributions among different populations in GCs is crucial to understand the origins of multiple populations. 

However, directly measuring He abundance of stars in GCs through spectroscopy is challenging. This is because He photospheric absorptions cannot be excited in the atmospheres of late-type stars, and the measurement of He abundance in stars with atmospheric temperatures exceeding 11,000 K is affected by the Grundahl jump effect \citep{Grun99a}, direct measurements of He abundance in GCs are very limited. For cool stars, typically of FGK type, only the chromospheric He I 10830\AA\; spectral line can be used for He abundance determination \citep[e.g.,][]{Dupr13a}. However, a recent study presented by \cite{Jian24a} has revealed that the strength of this absorption line strongly correlates with the Ca II $\log{R'_{\rm HK}}$ index, indicating that the structure of the stellar chromosphere plays a significant role in the He I 10830\AA\; line. This makes it inaccurate to characterize stellar He abundance solely using the He I 10830 \AA\; absorption. Consequently, in GCs, only horizontal branch (HB) stars within a specific temperature range can be employed for the determination of He abundance \citep[e.g.,][]{Grat13a,Mari14a}. This makes it challenging to obtain a complete understanding of He abundance distribution among stars in GCs through spectroscopy. First, not all GCs have HB stars in this specific temperature range. Second, in GCs with MPs, the HB stars from the first and second generations differ significantly in color, making it difficult to measure the He abundance of HB stars for both populations. Because of this, in addition to direct spectroscopic observations, He variations among GCs are explored through photometric studies, such as extended or multiple HBs \citep[e.g.,][]{Dale13a,Milo18a}, extended or dual red giant branch (RGB) bumps \citep{Nata11a}, extended or multiple main sequences \citep[MS,][]{Norr04a,Piot07a,Cade23a} and their combinations \citep[][]{Bell13a,Nard15a}. Studies on helium abundance in clusters within satellite galaxies have only recently increased due to their great distance. \cite{Lagi19a} reported evidence of He enhancement ($\delta{Y}\sim0.01$) in the second population stars from four GCs in the Small Magellanic Cloud (SMC). \cite{Chan19a} found that four of the eight clusters in the Magellanic Clouds show indications of helium variation based on their HB morphologies. Our previous study has shown that the Large Magellanic Cloud (LMC) cluster NGC 2210 shows evidence of He variations of approximately $\delta{Y}$=0.06$-$0.07 among its MS dwarfs \citep{Li23a}. In this study, we found that this research underestimated the impact of photometric errors in the cluster's central region. We have corrected this value to at least ($\delta{Y} $=0.08), but this adjustment applies only to the cluster members of NGC 2210 outside the central area. Thanks to the Multi Unit Spectroscopic Explorer (MUSE) integral-field spectrograph based on the Very Large Telescope (VLT), it subsequently became feasible to perform spectral measurements of the MS stars within MC clusters, although measuring their He abundance can currently be made only for the early-type MS stars in young MC clusters. \cite{Cari20a} estimated the He abundance of ten bright MS stars in the SMC cluster NGC 330, and they found no evidence of variation in helium abundance among the individual stars. They subsequently analyzed another pair of double clusters in the LMC, NGC 1850A and NGC 1850B, and discovered significant differences in He abundance among the B-type main-sequence stars, with those belonging to NGC 1850B exhibiting a helium abundance exceeding that of NGC 1850A by more than \citep[$\delta{Y}$=0.1,][]{Cari24a}. 

Using the morphology of the HB to fit the He distribution of stellar populations is the most common method, as HB stars are very bright and their morphology often shows a wide range in color and magnitude, which helps reduce uncertainties caused by photometric errors. However, because HB is a very advanced phase of stellar evolution, models that rely on the analysis of HB morphology are usually subject to larger systematic errors. It is well known that the He variation is not the only parameter governing the HB morphology. Multiple factors may work together to influence the morphology of the HB, including the metallicity of stellar populations, age dispersion, and mass loss rates. Especially in GCs that contain MPs, HB stars from different stellar generations may undergo significantly different mass loss histories \citep{Tail20a}. This further complicates the ability of the HB morphology to accurately represent the differences in their He abundances. Using the RGB bump to study the He abundance distribution in star clusters results in lower uncertainty due to the influence of stellar mass loss \citep[e.g.,][]{Lagi18a}. However, this method is still affected by the metallicity of the stellar population. Moreover, even with the reasonable assumption that the metallicity of stars in the cluster is uniform, unresolved binary stars can substantially alter the luminosity function distribution of the RGB, thus obscuring the morphological features of the RGB bump \citep[e.g.,][]{Li23a}. Importantly, the RGB bump stars in a cluster typically make up a small proportion of the total population, making it nearly impossible to detect a clear RGB bump structure in GCs that lack a significant number of red giant stars.

Compared to the methods mentioned above, examining the morphology of the MS of GCs to find their He abundance distribution is more physically reliable. As helium is the second most abundant element in stars, variations in its abundance have a significant impact on stellar structure and evolution. First, the initial helium abundance in MS stars of a specific mass affects their effective temperature ($T_{\rm eff}$) because the associated decrease in the radiative opacities of their outer layers makes them hotter as the He abundance ($Y$) increases. Second, an increase in the $Y$ raises the molecular weight of the stellar gas, causing MS stars of a specified mass also to exhibit greater luminosity. Moreover, stars that are rich in helium evolve more quickly than normal stars, as they possess higher luminosities. The overall effect is that in the MS region below the MS turn-off (MSTO), He-enriched MS stars will appear significantly bluer compared to normal stars. As they approach the MSTO, the branch of He-enriched stars will gradually come closer to the normal MS, eventually connecting with the red end of the MSTO region and forming a dimmer subgiant branch (SGB). MS stars have not experienced significant mass loss or convective dredge-up, making their He abundance a more reliable indicator of the primordial He distribution of the GC. A series of studies have demonstrated that with the correct selection of optical bands, the multiple MSs seen in clusters can accurately reflect the internal He variations \citep[e.g.,][]{Sbor11a, Cass13a}. Therefore, even though the luminosity of the MS is considerably less than that of the RGB bump and the HB, examining the He distribution in clusters by analyzing the morphology of the MS is a worthwhile method, especially when complemented by deep exposure observations, particularly from space telescopes. 

To measure the He abundance of stellar populations through the morphology of the MS, the selected photometric bands must meet the following conditions: 1. The color should be sensitive to variations in helium abundance (i.e., $dY/d{\rm color}$). 2. The color should not be sensitive to changes in other light elements due to the presence of MPs. 3. The relationship of $dY/d{\rm color}$ should account for different physical inputs (such as adopted mixing length $\alpha_{\rm MLT}$, and/or atomic diffusion efficiency). A detailed theoretical study has shown that the F606W and F814W bands of the {\sl Hubble} Space Telescope's Advanced Camera Survey ({HST}/ACS) effectively satisfy the requirements for photometric passbands. The color distribution (F606W $-$ F814W) of the MS obtained from these bands can reliably represent the He abundance distribution within stellar populations \citep{Cass17a}. 

In this study, we analyzed the MS morphology of four GCs in the LMC, Hodge 11, NGC 1841, NGC 2210, and NGC 2257, with the aim of determining the He abundance distribution among their member stars. Our analysis relies on color-magnitude diagrams (CMDs) utilizing the F606W and F814W passbands of the {HST}/ACS. We observed a significant broadening of the MS in all four clusters, which, after minimizing the effects of field star contamination, photometric errors, unresolved binaries, differential extinction, and variations in metallicity, can only be explained by He abundance variation. We infer that there are likely significant variations in He abundance of at least $\delta{Y}$ = 0.08 within these clusters. 

The following sections of this paper will include the subsequent topics: Section \ref{S2} will describe our methods for data reduction and our analysis, including photometry, cluster members determination, isochrones fitting, and other statistical analysis using ASs. Section \ref{S3} presents our main results, primarily focused on the He distribution of MS stars within the clusters, including the values of He enrichment ($\delta{Y}$) of MS stars and the fraction of He-enriched stars. Section \ref{S4} contains a scientific discussion addressing these results along with our conclusions.

\section{Data reduction and methods}\label{S2}
\subsection{Calculation of structural parameters}
\begin{figure}[ht!]
\begin{center}
\includegraphics[width=0.5\textwidth]{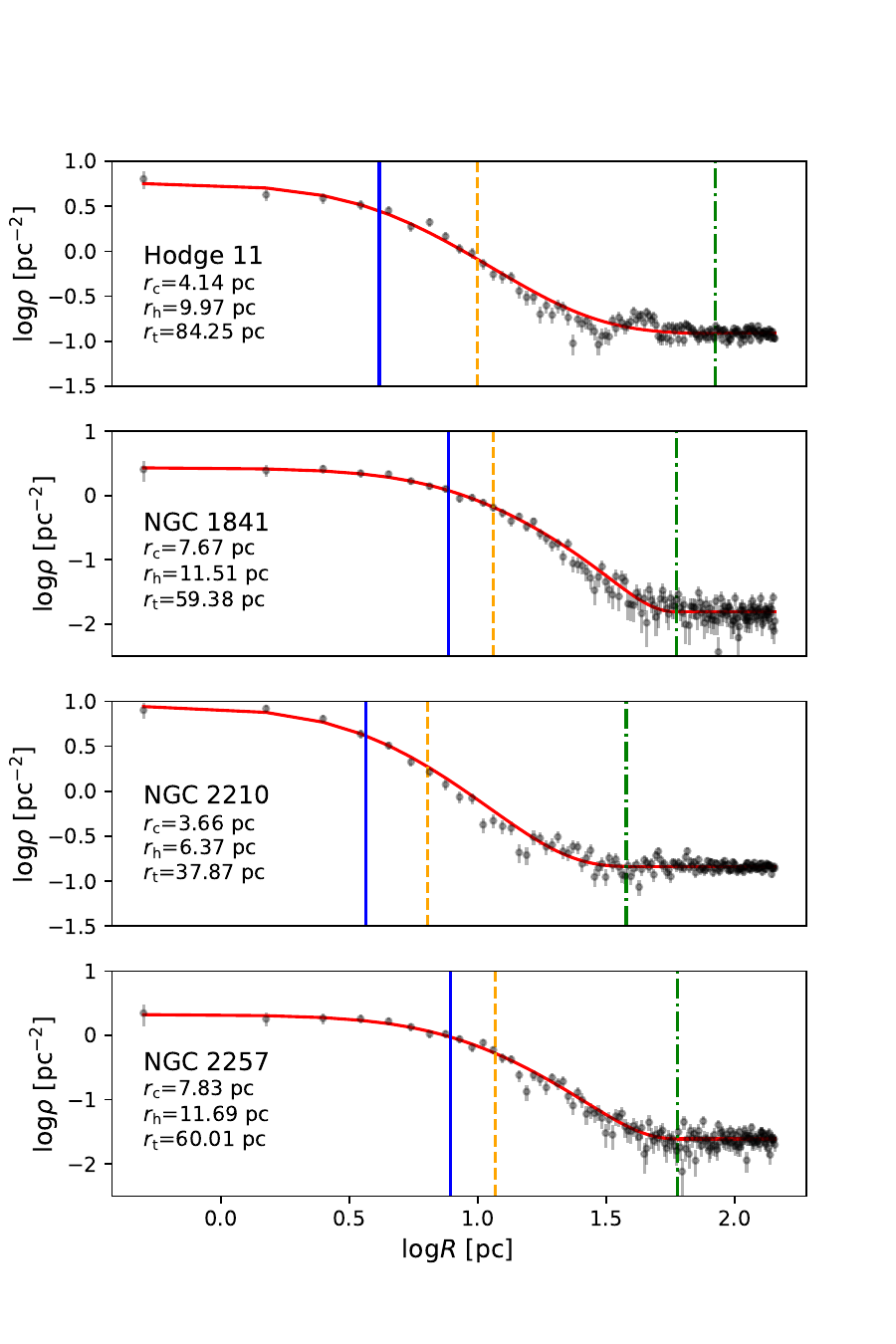}
  \caption{Stellar Number Density Profiles of Hodge 11, NGC 1841, NGC 2210, and NGC 2257. 
Best-fitting King profile shown by solid red curves. Solid blue line, dashed orange line, and dash-dotted green line indicate core radius, half-number radius, and tidal radius, respectively.}
  \label{F1}
\end{center}
\end{figure}

Our analyses are on four GCs within the LMC, Hodge 11, NGC 1841, NGC 2210, and NGC 2257. We first collected the spatial distribution of all stars with $G\leq$20 mag within 10 arcminutes of the centers\footnote{\rm resolved by Sesame Strasbourg (Simbad-NED-VizieR)} of these four clusters ($\sim$144 pc at the distance of the LMC), utilizing the archival dataset from the Gaia DR3 \citep{Gaia23a}. We employed these catalogs to determine their structural parameters by fitting the radial profiles of their number density with a King profile \citep{King62a}: 
\begin{equation}
\rho(r)=k\left[\frac{1}{\sqrt{1+\left(r / r_{\mathrm{c}}\right)^2}}-\frac{1}{\sqrt{1+\left(r_{\mathrm{t}} / r_{\mathrm{c}}\right)^2}}\right]^2+b,
\label{eq1}
\end{equation}
in this equation, the core and tidal radii are referred to as $r_{\rm c}$ and $r_{\rm t}$, respectively. The symbol $b$ represents the number density of the background field, while $k$ is a normalization coefficient. Additionally, $\rho$ stands for the number density, and $r$ indicates the distance from a star to the center of the cluster. The best-fitting King profiles are employed to calculate the half-number radii, referred to as $r_{\rm h}$. {For old GCs, the majority of stars are low-mass stars, which contribute most of the cluster's total mass. Therefore, the half-number radius can generally be considered equivalent to the half-mass radius.} The derived {\rm $b$} are subsequently used for field decontamination. In Figure \ref{F1} we show the number density profiles of our clusters as well as the best fitting King profile. The derived structure parameters are summarized in Table \ref{T1}. 

\begin{table*}
\caption{Derived structure parameters of GCs. }
\label{T1}
\centering
\begin{tabular}{cccccc}
\hline\hline
Cluster name & $\alpha_{\rm J2016}^{(1)}$ & ${\delta_{\rm J2016}}^{(1)}$ & $r_{\rm c}^{(2)}$ (pc) & $r_{\rm h}^{(3)}$ (pc) & $r_{\rm t}^{(4)}$ (pc)  \\
\hline
Hodge 11 & $06^{\rm h}14^{\rm m}22.92^{\rm s}$ & $-69^{\circ}50^{\rm m}54.90^{\rm s}$ & 4.14$\pm$0.16 & 8.25$_{-0.26}^{+1.52}$ & 84.25$\pm$32.33  \\
NGC 1841 & $04^{\rm h}45^{\rm m}24.37^{\rm s}$ & $-83^{\circ}59^{\rm m}53.30^{\rm s}$ & 7.60$\pm$0.16 & 11.23$_{-0.18}^{+0.14}$ & 59.38$\pm$3.79  \\
NGC 2210 & $06^{\rm h}11^{\rm m}31.29^{\rm s}$ & $-69^{\circ}07^{\rm m}17.04^{\rm s}$ & 3.66$\pm$0.09 & 6.13$_{-0.14}^{+0.20}$ & 37.87$\pm$3.59  \\
NGC 2257 &  $06^{\rm h}30^{\rm m}12.60^{\rm s}$ & $-64^{\circ}19^{\rm m}40.01^{\rm s}$ & 7.83$\pm$0.20 & 11.39$_{-0.24}^{+0.16}$ & 60.01$\pm$4.76  
\\
\hline
\end{tabular}
\\
\tablefoot{(1) Center coordinate; (2) Core radius; (3) Half-number radius; (4) Tidal radius.}
\end{table*}

We compared our results with those of \cite{Lanz19a} and found that, in terms of $r_{\rm c}$ and $r_{\rm h}$ (in their results, $r_{\rm h}$ means half-mass radius), our findings are consistent with their trends. For $r_{\rm c}$, the size ranking of the four GCs is NGC 2210$<$Hodge 11$<$NGC 1841$\approx$NGC 2257. Similarly, for $r_{\rm h}$, the size ranking again remains NGC 2210$<$Hodge 11$<$NGC 1841$\approx$NGC 2257. Within the margin of error, our $r_{\rm c}$ are systematically larger than those of \cite{Lanz19a}. In contrast, for $r_{\rm h}$, our results are consistently larger than theirs. Notably, our analysis suggests that Hodge 11 may possess the largest $r_{\rm t}$ among the four clusters, whereas their fitting placed Hodge 11 third in the ranking. However, our fitting results for Hodge 11 exhibit the largest errors among the four clusters, which may be attributed to significant fluctuations in the stellar density surrounding Hodge 11 (see the top panel of Figure \ref{F1}). Given that we utilized different datasets for our fitting and that our analyzed field of view (FoV) is larger than theirs (their fittings are based on the {HST} ACS/WFC data), the observed discrepancies are acceptable. Therefore, we conclude that our fitting is generally in agreement with the findings of \cite{Lanz19a}. 

\subsection{Photometry}
Our analysis mainly used data obtained from {HST} ACS/WFC observations because it has a deeper exposure than the Gaia DR3. The program ID is GO-14164 (PI: Sarajedini). To improve the accuracy of removing field star contamination, we also used the observational data from the Ultraviolet and Visible Channel of the Wide Field Camera 3 (UVIS/WFC3) on the {HST} for the four clusters in a similar F814W passband. This program ID is GO-16748 (PI: Niederhofer). The average timeline between the two observations for our four clusters is $\sim$6 yr. We will subsequently employ both datasets to obtain the relative proper motions (PMs) of the stars within their overlapping FoV, thereby facilitating a more precise field star decontamination. The basic information about the dataset we used is summarized in Table \ref{T2}. 

For each cluster, we use the specific {HST} photometric package {\sc Dolphot2.0} \citep{Dolp11a,Dolp11b,Dolp13a} to conduct point-spread function (PSF) based photometry on their charge-transfer-efficiency-corrected frames (the images with root names ending in `\_flc' and `\_drc'). The photometric procedure we employ involves the following steps. First, we identify and mask bad pixels in the images. This step is automatically carried out using the {\tt acsmask} command for ACS/WFC data and the {\tt wfc3mask} command for UVIS/WFC3 data. Next, we separate the charge-transfer-efficiency-corrected frames into two chips using the {\tt splitgroups} command. We then compute the corresponding sky images for all frames in order to estimate the background levels at different locations, utilizing the {\tt calcsky} command. In this step, we set the parameters based on the tutorial's recommendations, choosing {\tt step}=4 in the sky-fitting settings. At this point, the program will fit the sky within the photometry aperture as a four-parameter PSF fit, which is the most accurate sky calculation provided by the photometric package. After completing the previous steps, we begin photometry using the {\tt dolphot} command. At this stage, we will choose the drizzled image (`*\_drc.fits', per filter) as the reference frame. Since we will attempt to reduce field star contamination by using the changes in star positions, PM, from two observations, we need to eliminate errors caused by factors such as CCD distortion and alignment as much as possible. We use the WCS header information for alignment, setting {\tt UseWCS}=2 to estimate a full distortion solution.

The original star catalog we obtained contains a large amount of information, including a total catalog formed by merging all frames and individual catalogs from each separate frame. Since we are analyzing MS stars, short exposure times can lead to high photometric errors, making it impossible to estimate the distribution of He abundance within the cluster. Therefore, we decide to choose the combined catalog from all frames as our analysis sample. We first remove false detections identified by {\sc Dolphot}; these detections will have a magnitude of 99.999 mag in at least one passband. We then carried out a quality filter on the valid detections, choosing only the stars clearly labeled as good stars (Object types=1). We removed sources with sharpness values less than $-$0.3 and greater than 0.3. According to the {\sc Dolphot} manual, these sources are more likely to be cosmic rays ($>$0.3) or extended sources (such as blends or galaxies, $<-$0.3). Roundness describes the ellipticity of a source's light profile. A perfectly circular source has a Roundness value of 0. In this study, we exclusively select sources with a Roundness value lower than 0.5, as sources exhibiting higher roundness are unlikely to be perfect stars (e.g., those blended with diffractive spikes). Crowding refers to the level of contamination caused by nearby sources. The Crowding value, expressed in magnitudes, indicates how much brighter the star would have appeared if the adjacent stars had not been fitted simultaneously. We applied a fairly relaxed limitation on crowding, excluding only sources with a crowding value greater than 0.5 mag. We discovered that most of the MS stars we are interested in had significant crowding, as their intrinsic brightness was similar to the background level. Finally, we also removed sources with an SNR$<$5, as these sources had too large errors that could significantly affect the analysis. We will then analyze artificial stars (ASs) that went through the same selection criteria to statistically compare whether the morphology of the MS matches the predictions of the simple-stellar population (SSP) models.

\subsection{Estimation of stellar proper motions}\label{PM}

The header files of the images used for photometry contain World Coordinate System (WCS) keywords, which define the relationship between pixel coordinates in the image and celestial coordinates. Specifically, these files include eight parameters, among which {\tt CRPIX1} and {\tt CRPIX2} represent the pixel coordinates of the reference point to which the projection and rotation pertain. {\tt CRVAL1} and {\tt CRVAL2} specify the central coordinates in terms of right ascension (RA., or $\alpha$) and declination (DEC., or $\delta$), or longitude and latitude, expressed in decimal degrees. The FITS WCS standard employs a rotation matrix consisting of {\tt CD1\_1}, {\tt CD1\_2}, {\tt CD2\_1}, and {\tt CD2\_2} to indicate both rotation and scale, facilitating more intuitive computation in cases where the axes are skewed. In particular, the conversion of the X and Y coordinates on the CCD to RA and DEC can be done using the following relationships:

\begin{equation*}
{\delta} = {\tt CRVAL2}+\left[{\tt CD2\_1}(X-{\tt CRPIX1})+{\tt CD2\_2(Y-{\tt CRPIX2})}\right]
\end{equation*}
\begin{equation*}
{\alpha} = {\tt CRVAL1}+\frac{\left[{\tt CD1\_1}(X-{\tt CRPIX1})+{\tt CD1\_2(Y-{\tt CRPIX2})}\right]}{\cos{{\delta}}}
\end{equation*}

To more accurately identify the member stars of the star cluster, we used a ``fool man's method'' to assess the PMs of each star using datasets with a time interval of about 6 years. For each star observed by ACS/WFC around $\sim$2016 (first observation), we looked for its matching observation in the UVIS/WFC3 data from $\sim$2022 (second observation) by pairing it with the source that was nearest in distance during the second observation. For each star, we recorded the distance to the nearest star identified in the $\sim$2022 observation. We found that the distribution of the nearest distances showed three clear different peaks. The first peak represents stars that were not covered by the FoV of the second observation due to incomplete overlap between the two observational fields. In this situation, our method randomly selects a star from the second observation as its nearest counterpart, leading to very large nearest distances for these stars, which we then excluded first. The second peak mainly results from the different exposure depths of the two observations, which causes some stars to be missed in the second observation. In this situation, our method still randomly picks the nearest star from the second observation as its match; however, since both stars are now in the same FoV, the nearest distance is much smaller than in the first peak. Nevertheless, we can easily remove these stars by looking at their two-dimensional distribution in ($\Delta{\alpha}$) and ($\Delta{\delta}$). The third peak corresponds to nearest distances that are much smaller than those of the first two peaks, representing the same stars detected in both observations. From the distribution, this nearest distance is very close to zero, as the movement of the stars over six years is not enough to create noticeable changes in position. However, we still see that this peak has an extension towards larger distances, which we think comes from the influence of field stars. To more accurately distinguish between field stars and member stars, we further analyzed the distribution of these stars in a two-dimensional PM space ($\Delta{\alpha}\cos{\delta}$, $\Delta{\alpha})$). We found that their overall distribution remains relatively continuous, and it does not exhibit an isotropic uniform distribution. This indicates that relying solely on a 6 yr time baseline results in low accuracy for measuring PM in terms of direction. We then created iso-density contour maps of these stars based on their distribution in the two-dimensional PM diagram. Using the previously calculated field star contamination rate, we chose specific contour lines as boundaries. The stars inside these contour lines are classified as member stars of the star cluster, while those outside are considered field stars. 

\begin{figure*}[ht!]
\begin{center}
\includegraphics[width=1.0\textwidth]{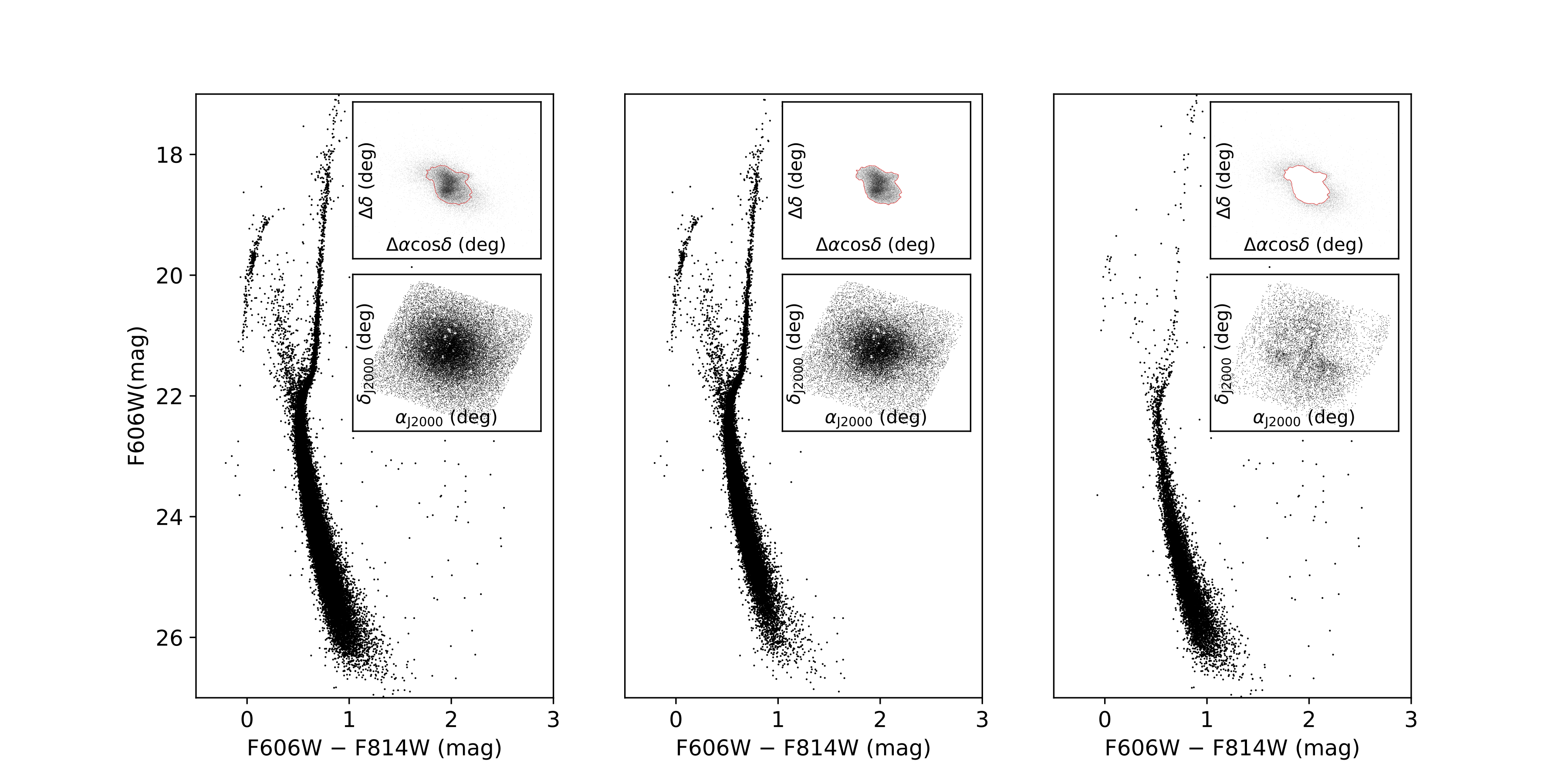}
  \caption{CMD of Hodge 11. Left: CMD shows all stars observed in the field of cluster Hodge 11. Middle: CMD displays stars classified as cluster members. Right: CMD features stars identified as field stars. In each panel, inset plots illustrate differences in coordinates between two observations taken approximately 7 years apart (top) and the spatial distribution of the corresponding stars.}
  \label{F2}
\end{center}
\end{figure*}

The PM distribution of these stars and the contour lines we selected are displayed in the subplots of Figure \ref{F2}, and figures \ref{F3}--\ref{F5}. These four figures show the CMDs of all observed stars in four clusters (left panel), the CMDs of the selected cluster member stars (middle panel), and the CMDs of the identified field stars (right panel). In the upper right corner of each diagram, we provide two subplots: the PM distribution of the corresponding stars (top) and their spatial distribution (bottom). We found that, based on the contamination rate we calculated and the relative PM distribution from two observations, we could effectively remove most field stars. Specifically, the CMDs of the removed field stars are quite scattered, clearly deviating from the isochrones that describe the star cluster. Additionally, the spatial distribution of the removed field stars is mostly uniform, showing no clear clustering, indicating that we did not overestimate field star contamination\footnote{We also conducted an analysis of the same cluster using the recently released catalog by \cite{Nied24a}. We found that their derived distribution of stellar PMs is relatively more uniform (with the two-dimensional distribution tending towards an elliptical shape); however, the larger errors result in a nearly complete mixing of field stars and member stars. Specifically, using their catalog, when we employed iso-density contours that meet the same criteria to filter member stars, we observed that the discarded field stars still exhibited a clear clustered distribution. Even when we significantly relaxed the criteria for the contour lines (allowing member stars to have very large PM), we still inadvertently over-subtracted an excessive number of cluster member stars.} (Except for NGC 2210, we found that no matter how we adjusted the criteria for removing field stars, the eliminated field stars still showed some clustering. In the future, with updated observational data (allowing for PM calculations based on a longer age baseline), we may be able to improve this situation). 

\subsection{Artificial stars}
{\sc Dolphot} allows for the creation of ASs on the original images, based on the PSFs used in the initial photometry. Before this, we need to set the magnitudes of the ASs. To make the ASs as realistic as possible, we first fit isochrones to the observed star cluster. The isochrones we used are derived from the Princeton-Goddard-PUC (PGPUC) stellar evolutionary code\footnote{\url {http://www2.astro.puc.cl/pgpuc/index.php}} \citep{Valc12a,Valc13a}. We used this model because it is specifically made for studying low-mass dwarfs, which perfectly fits our research goals. The inputs for the PGPUC model comprise the values of age, He mass abundance (Y), global metallicity (Z), solar-scaled abundance of $\alpha$-element ([$\alpha$/Fe]), and mass-loss rate ($\eta$). Among the parameters mentioned, age, $Y$, $Z$,  adopted distance modulus, $(m-M)_0$, and extinction, $E(B-V)$, have the most substantial impact on our fitting, while [$\alpha$/Fe] and $\eta$ exert minimal influence. For all star clusters, we have consistent inputs of [$\alpha$/Fe]=+0.3 \citep{Wagn17a} and $\eta$=0.2 (default values). The best-fitting isochrones are determined through visual inspection. As a result, our estimation of the errors in the fitting parameters is rather informal, simply matching the grid density used during our fitting. Furthermore, $Y$=0.25 is always kept fixed in our fitting, as it serves as a reference for creating MPs with different $Y$ values. The fitting parameters are summarized in Table \ref{T3}. Our fitting results are consistent with those of \cite{Wagn17a} within the margin of error.

\begin{table*}
\caption{Global parameters derived from isochrones fitting}
\label{T3}
\centering
\begin{tabular}{cccccc}
\hline\hline
Cluster & Age (Gyr) & $Y$ & $Z$ & $(m-M)_0$ (mag) & $E(B-V)$ (mag) \\
\hline
Hodge 11 & 13.5$\pm0.6$ & 0.25 & 0.0005$\pm0.0001$ & 18.58$\pm0.05$ & 0.05$\pm0.01$ \\
NGC 1841 & 13.0$^{+0.5}_{-0.7}$ & 0.25 & 0.0003$\pm0.0001$ & 18.40$\pm0.05$ & 0.17$\pm0.01$ \\
NGC 2210 & 12.5$^{+0.7}_{-0.8}$ & 0.25 & 0.0002$\pm0.0001$ & 18.40$\pm0.05$ & 0.06$\pm0.01$ \\
NGC 2257 & 12.5$^{+0.7}_{-0.8}$ & 0.25 & 0.0004$\pm0.0001$ & 18.40$\pm0.05$ & 0.07$\pm0.01$ \\
\hline
\end{tabular}
\tablefoot{$Y$ is fixed at 0.25.}
\end{table*}

Clearly, using visual inspection for isochrone fitting introduces considerable subjective bias. To create synthetic stellar populations that closely match real observations, we use the MS ridgeline (MSRL) as a reference for generating ASs. Similar to \cite{Li23a}, we determined the MSRL below the TO region using the robust regression method based on the Gaussian process and iterative trimming created by \cite{Li20a,Li21a}. The PGPUC isochrones in our analysis mainly aim to show how the colors of stellar populations change with varying $Y$, i.e., $\Delta{Y}/\Delta{({\rm F606W - F814W})}$. We subsequently calculated 11 isochrones with varying He enrichments of $\Delta{Y}$ = 0.01 to 0.12 ($Y$ = 0.26 to 0.37, with an increment of 0.01). Thus, for each star cluster, we calculated 13 synthetic stellar populations with different values of $Y$ (including the fitted value of $Y$ = 0.25). The color deviations for each isochrone relative to the standard isochrone ($Y$ = 0.25), represented as $\Delta{({\rm F606W - F814W})}$, are then incorporated into the MSRL. 

We then used {\sc Dolphot} to generate ASs. The magnitudes of the ASs are derived from the MSRL, while their luminosity function is derived from the standard Kroupa mass function. Each population consists of 2 million ASs. Therefore, for each cluster, we generated 2.6 $\times$ 10$^7$ ASs with $Y$ values from 0.25 to 0.37. Since ASs are created using the same PSF as real stars, and we have filtered them as we did for real stars, they can, in principle, simulate an observation with similar photometric uncertainties (including noise from cosmic rays, hot pixels, crowding effects, and other artifacts) as real stars. For each synthetic population, we added additional color noise to simulate the residuals after the removal of differential reddening. The differences between the recovered and input magnitudes of ASs are referred to as the photometric uncertainties\footnote{The uncertainties returned by {\sc Dolphot} photometry are based on the measured flux and represent instrumental uncertainties, which are often significantly smaller than the actual photometric errors.}. 

Based on the photometric errors derived from ASs, we decided to analyze MS stars within the magnitude range of 22.5 mag$<$F606W$<$ 24.5 mag. MS stars brighter than this magnitude range are too close to the MSTO region, where increased helium abundance accelerates stellar evolution, which offsets the effects of luminosity and temperature changes caused by He enrichment. Below this magnitude range, MS stars have too low SNRs, leading to the spread of the MS being dominated by photometric uncertainties, which would mask any possible signals of He variation. For each cluster, ASs indicate an average completeness of $\geq$65\% for MS stars within this magnitude range. 

Through ASs, we also found that, within $\sim$38 arcsec (about 9 pc) from the center of each cluster, the photometric uncertainties increase sharply due to crowding. Therefore, we decided to analyze only the sample of MS stars located between 10 pc and 30 pc. Beyond 30 pc, the proximity of stars to the edge of the FoV also leads to a slight increase in both photometric and astrometric uncertainties, see Figure \ref{A1}.

\subsection{Differential extinction}\label{reddening}

The dust distribution in the ACS/WFC FoV (202$^{\prime\prime}$× 202$^{\prime\prime}$, corresponding to 48.5 pc × 48.5 pc at the distance of the LMC) may be inhomogeneous. We have used the method outlined in \citet{Milo12} to minimize the effect caused by the dust inhomogeneity--the differential reddening. Briefly, to quantify the amount of differential reddening suffered by each star across the FoV, we chose RGB and SGB stars instead of MS stars as the reference population to avoid underestimating the helium spread. We derived the fiducial line of the reference population based on a 2D probability density function in the F606W versus F606W$-$F814W CMD. To simplify the correction process, we rotated the CMD counterclockwise to define a new reference frame in which the abscissa (${\rm X'}$) and the ordinate (${\rm Y'}$) are parallel and orthogonal to the reddening direction, respectively. For each star, we calculated the average systematic offset ($\overline{\Delta {\rm X'}}$) from the fiducial line, based on the nearest 50 reference stars, excluding the target star from the computation. To correct the photometry for the effects of differential reddening, we subtracted the derived $\overline{\Delta {\rm X'}}$ from the ${\rm X'}$ value of each star. After applying this correction, the de-reddened CMD was used to refine the selection of reference stars and update the fiducial line. This process was iterated until convergence, typically after about three iterations. The corrected ${\rm X'}$ and ${\rm Y'}$ were transformed back into F606W and F814W magnitudes, using the relative absorption coefficients from the extinction curves of \citet{Card89} and \citet{Odon94}.

We constructed a series of extinction distributions utilizing ASs (as detailed in the following subsection) and then calculated the extinction of the ASs using the same method employed for real stars. By comparing the calculated extinction to the input values, we estimate the extinction residuals associated with our approach. Specifically, we randomly selected ten points within the FoV of the ASs, assuming their extinctions were randomly drawn from a distribution function with varying dispersion. We then performed a two-dimensional polynomial fitting of extinction based on these ten points across the entire FoV. The extinction for other ASs in the field was obtained by interpolating this fit, resulting in a specific differential extinction distribution. We applied the input extinction to correct the input magnitudes of each AS by using the same method applied to the real observation, and obtained the photometric magnitudes of the ASs. We used the above statistical methods to calculate the differential extinction for each AS, and the calculated differential extinction was used to correct the photometric magnitudes (not input magnitudes) of the ASs. We also performed photometry on the ASs that did not include the differential extinction. By comparing the ASs that included differential extinction with those that did not, we obtained the color dispersion difference in their MS, $\delta{\rm (F606W-F814W)}$, which we conclude describes the extinction residuals after de-reddening. 

In Figure \ref{A2}, we present an example to demonstrate the accuracy of our de-extinction method. The upper left panel shows a spatial distribution of an AS population, with color representing their input extinction. The upper right shows the same distribution but with color representing their calculated extinction. We found that the calculated extinction distribution matches the input distribution well. The bottom panel shows the correlation between the input differential extinction and the color residuals after correcting for differential extinction. We found that, except for one outlier, there is a positive correlation between the input differential extinction and the color residuals. We use a linear function to fit this positive correlation and make corrections to the subsequent analyses of the ASs.

\subsection{Quantify helium spread}

Our subsequent analysis is essentially the same as in \cite{Li23a}. For both the observed and simulated MSs, we calculated the color differences, $\Delta{\rm (F606W - F814W)}$, for each star relative to their MSRL. Here the simulated MSs can be composed of either a single population (i.e., an SSP with a fixed $Y$ value) or MPs with different $Y$ values. From each artificial stellar population, we select a subsample that matches both the luminosity function and the total number of stars of the real observations, serving as a representative sample. A synthetic MPs is then constructed as a composition of these synthetic populations. To minimize the influence of unresolved binaries, we selected only those MS stars located on the blue side of the MSRL for our analysis. For the sake of clarity, we "folded" the observed and simulated MS stars on the blue side of the MSRL, assuming that there exists a corresponding mirror star at a symmetrical position relative to the MSRL. We divided the observed and simulated MS stars into 76 bins based on their $\Delta{\rm (F606W-F814W)}$ values (where the number of bins is roughly equal to the square root of the number of observed stars) and counted the corresponding number of stars in each bin for both the observed and simulated data. We used the $\chi^2$ minimization method to evaluate the similarity between the observed and simulated distributions:
\begin{equation}
\chi^2=\sum_i\frac{(N^{\rm obs}_{\rm i}-N^{\rm mod}_{\rm i})}{N^{\rm obs}_{\rm i}}
\end{equation}
where $N^{\rm obs}_{\rm i}$ and $N^{\rm mod}_{\rm i}$ are the number of observed and simulated stars with their relative colors in the $i$-th bin.

Our model contains two parameters: one is $\delta{Y}$, defined as the difference between the maximum $Y$ in the synthetic stellar populations and $Y$ = 0.25 (the He abundance of the primordial stellar population). The other parameter, $f_{\rm 2P}$, represents the ratio of stars with $Y$ > 0.25 to the total number of stars in the synthetic population. Similar to \cite{Li23a}, we assume two extreme cases for the $Y$ distribution: one where $\delta{Y}$ is uniformly distributed, meaning that the number of stars with He enhancement is the same across different $Y$ values (for $Y$ > 0.25). For example, if we set $\delta{Y}$ = 0.05, the model stellar population would contain six populations with $Y$ values of 0.25, 0.26, 0.27, 0.28, 0.29, and 0.30. If we assume $f_{\rm 2P}$ = 0.5, then the population with $Y$ = 0.25 would represent 50\% of the total number of stars, while the five He-enriched populations with $Y$ values of 0.26, 0.27, 0.28, 0.29, and 0.30 would each makeup 10\% of the total stellar count. The other case assumes a highly discrete distribution of $\delta{Y}$, where the model MPs consist of only two populations, corresponding to $Y$ = 0.25 and $Y$ = $Y_{\rm MAX}$. For example, in the case of $\delta{Y}$ = 0.05, the model would include only the populations with $Y$ = 0.25 and $Y$ = 0.30. If $f_{\rm 2P}$ = 0.5, this means that both groups represent 50\% of the total number of stars. The two scenarios described above represent two extreme distribution cases for a given $f_{\rm 2P}$. Therefore, the actual distribution of $\delta{Y}$ is likely to lie somewhere between these two cases.

\section{Main results}\label{S3}

\begin{figure}[ht!]
\begin{center}
\includegraphics[width=0.5\textwidth]{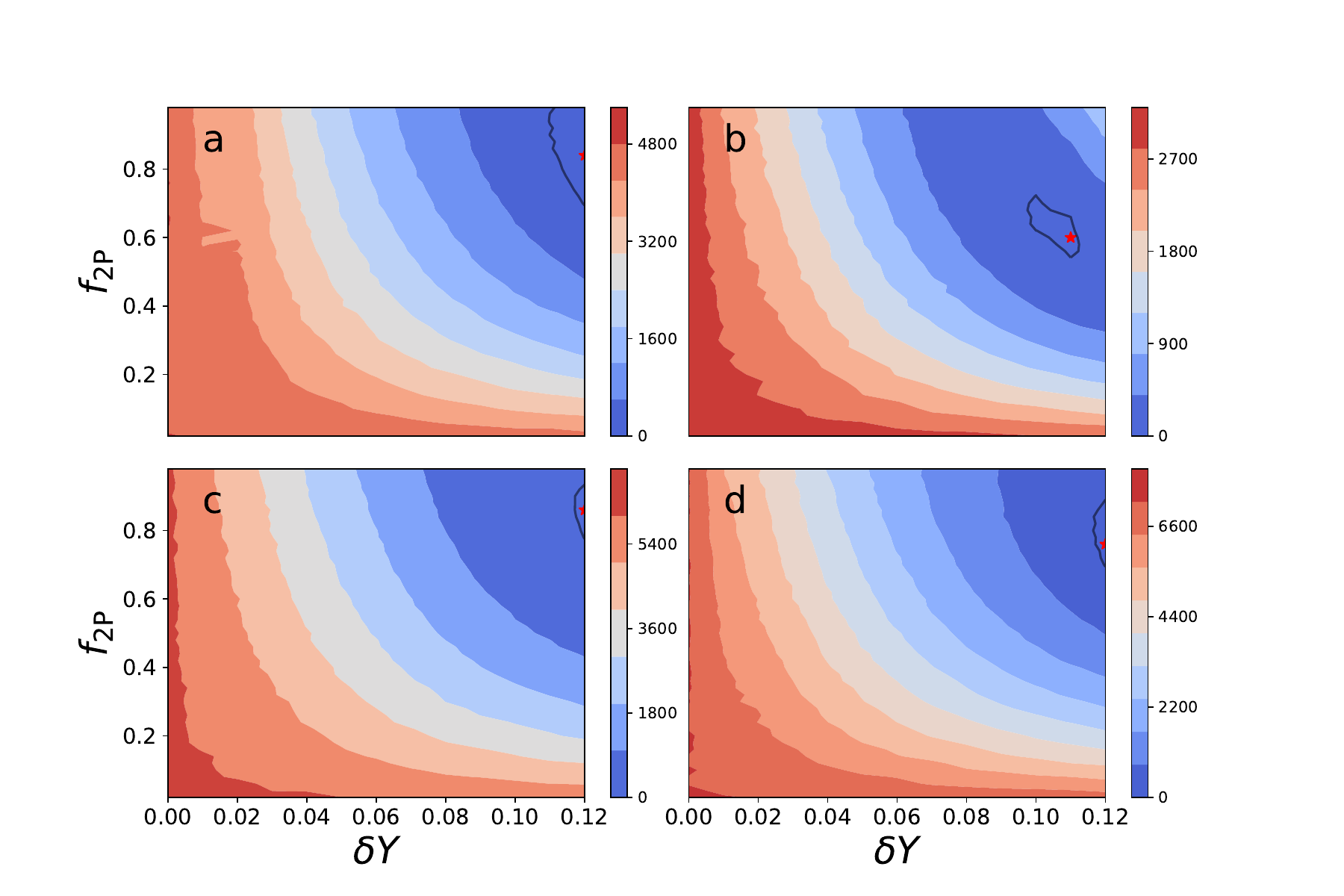}
  \caption{$\chi^2$ contour maps as a function of internal He spread, $\delta{Y}$, and fraction of He-enriched stars, $f_{\rm 2P}$. $\chi^2$ values are coded by colors. Best-fitting points (minimum $\chi^2$) are represented by red stars. Black solid lines indicate 1$\sigma$ uncertainties, calculated from $\chi^2$ variations. The adopted model assumes a uniform He distribution. Panels a, b, c, and d correspond to clusters Hodge 11, NGC 1841, NGC 2210, and NGC 2257, respectively.}
  \label{F6}
\end{center}
\end{figure}

\begin{figure}[ht!]
\begin{center}
\includegraphics[width=0.5\textwidth]{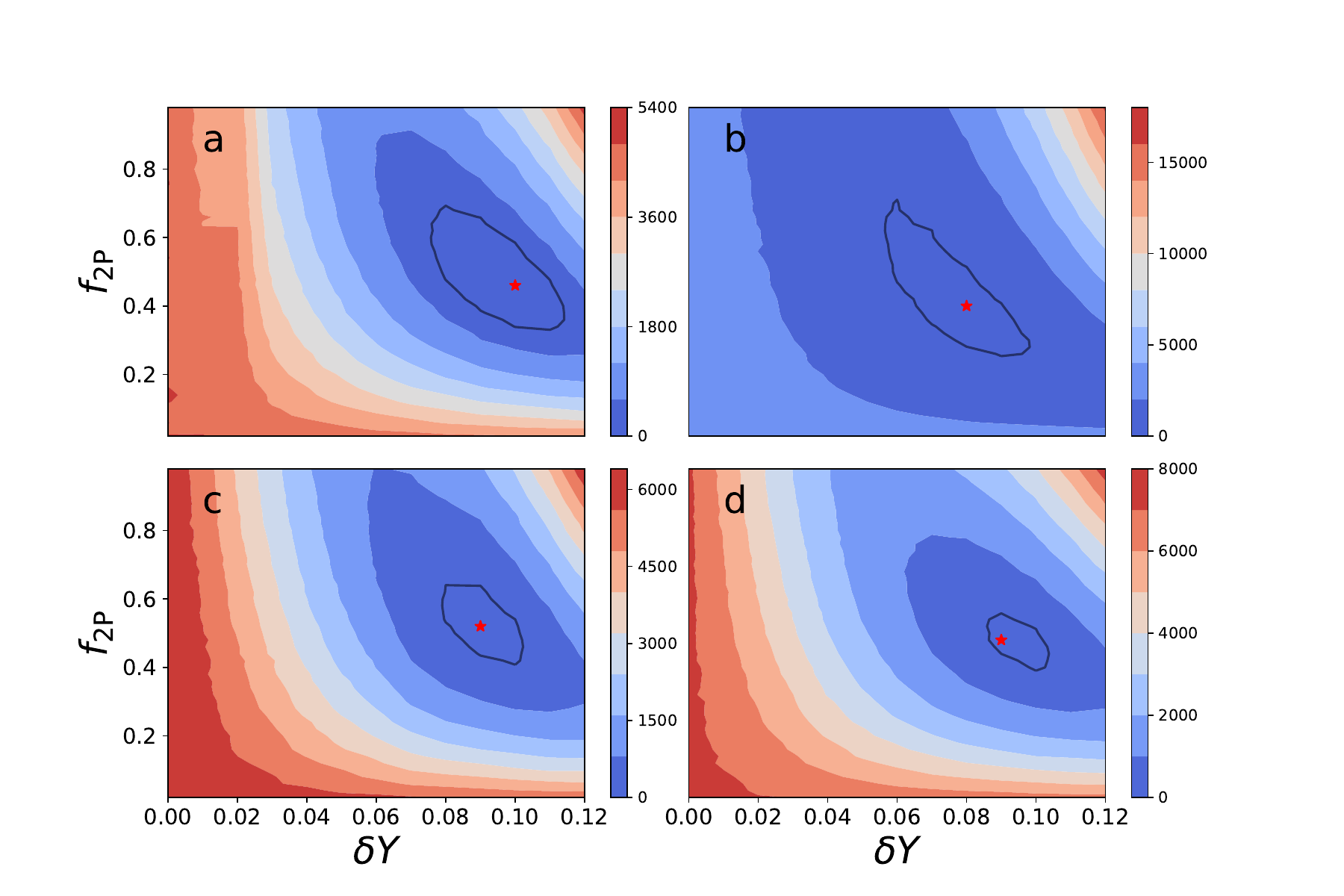}
  \caption{Same as in Fig. \ref{F6}, with the adopted model assumes a bimodal He distribution.}
  \label{F7}
\end{center}
\end{figure}

Our main results are shown in Figures \ref{F6} and \ref{F7}. The two contour figures show the distribution of $\chi^2$ obtained from comparing observations with models, with the value indicated by color. The x-axis represents the maximum He dispersion, $\delta{Y}$, while the y-axis shows the ratio of stars with He enrichment to the total population, $f_{\rm 2P}$. The point with the minimum $\chi^2$ is marked on the plot as a red pentagram. We fitted a third-order polynomial to the distribution of $\chi^2(\delta{Y}, f_{\rm 2P})$ and calculated the residuals $\delta{\chi^2}$. The contour line corresponding to the minimum $\chi^2$ + $\delta{\chi^2}$ is marked with a solid black line, indicating the fluctuation level of $\chi^2$ due to the random sampling effects of the model.

Figure \ref{F6} presents the results based on the assumption that $\delta{Y}$ follows a uniform distribution. We found that the derived best-fitting values for $\delta{Y}$ and $f_{\rm 2P}$ are substantially high for {all clusters}. With the exception of NGC 1841, our fitting indicates that both Hodge 11, NGC 2210, and NGC 2257 exhibit an internal helium dispersion of at least ($\delta{Y}$ = 0.12), with approximately 80\% of their stellar populations being He-enriched ($f_{\rm 2P}\sim0.8$. In comparison to the other three clusters, NGC 1841 exhibits a slightly lower internal He dispersion and fraction of He-enriched stars, with $\delta{Y}$=0.11 and $f_{\rm 2P}$=0.6. This implies that assuming a uniform distribution of helium, our fitting suggests that the stellar populations of all clusters are predominantly dominated by second-generation stars enriched in helium. 

If we assume a bimodal distribution of helium, our approach would yield a relatively moderate He-enrichment among the stellar populations, which is present in figure \ref{F7}. In this case, we find that the best-fitting ($\delta{Y}$) values for the four clusters are between 0.08 and 0.10, and the fraction of He-enriched stars is about 40\% to 50\%. This means that the numbers of first-generation and second-generation stars are quite similar in these clusters. We present the details of our $\chi^2$ fitting results in Table \ref{T4}. 

\begin{table}
\caption{Best-fitting internal helium spread, $\delta{Y}$, and fraction of He-enriched stars, $f_{\rm 2P}$.}
\label{T4}
\centering
\begin{tabular}{ccc|cc}
\hline\hline
Cluster & $\delta{Y}$ & $f_{\rm 2P}$ & $\delta{Y}$ & $f_{\rm 2P}$ \\\hline
Hodge 11 & $\geq$0.12 & 84\% & 0.10 & 46\% \\
NGC 1841 & 0.11 & 60\% & 0.08 & 40\% \\
NGC 2210 & $\geq$0.12 & 86\% & 0.09 & 52\% \\
NGC 2257 & $\geq$0.12 & 76\% & 0.09 & 48\% \\
\hline
\hline
\end{tabular}
\tablefoot{The values in second and third columns are based on the assumption that the $\delta{Y}$ distribution is flat, while the fourth and fifth columns are based on the assumption that the $\delta{Y}$ distribution is bimodal.}
\end{table}

\begin{figure*}[ht!]
\begin{center}
\includegraphics[width=1.0\textwidth]{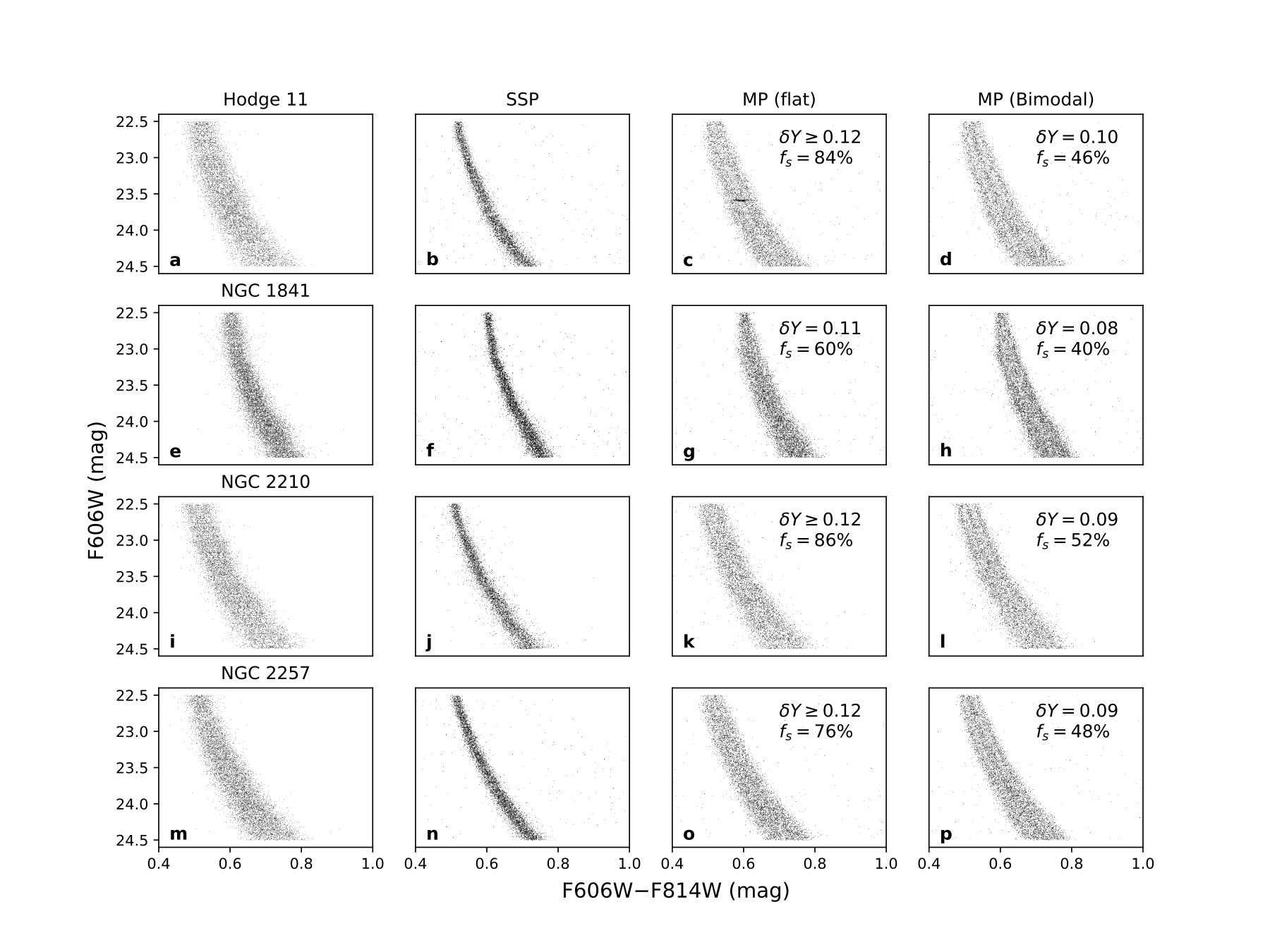}
  \caption{Comparison of MS observations and synthetic MS. From left to right are comparisons between the observed MS, and the synthetic MS of the SSP, of the best-fitting MPs with a uniform He distribution, and of the best-fitting MPs with a bimodal He distribution. From top to bottom are clusters Hodge 11, NGC 1841, NGC 2210 and NGC 2257.}
  \label{F8}
\end{center}
\end{figure*}

To guarantee the reasonableness of our results, we carefully selected several sets of best-fitting models and compared them with the observations through visual inspection. The results are illustrated in Figure \ref{F8}. We discovered that the observed widths of their MSs are indeed all significantly larger than what is predicted for an SSP. This broadening cannot be explained by photometric errors, differential extinction (including residuals after de-reddening), or unresolved binaries. Visual inspection indicates that the observed MS aligns more closely with the morphology of MPs containing helium dispersion. If we assume a uniform distribution of helium abundance, larger values of $\delta{Y}$ and $f_{\rm 2P}$ are required to reproduce the observed broadening of the MS. Conversely, if the distribution of helium abundance is bimodal, the observations can be reproduced with smaller values of $\delta{Y}$ and $f_{\rm 2P}$.

We employed the Modules for Experiments in Stellar Astrophysics \citep[MESA,][]{Paxt11a} to investigate whether our findings are consistent with the HB morphology. Given the extensive computational time necessary for stellar evolution calculations using MESA, we performed a qualitative analysis to determine the $\delta{Y}$ required for a morphological fit to the observations. Therefore, the stellar populations we examined all comprised a similar number of stars, with the $Y$ distribution ranging from 0.25 to 0.37, with increments of 0.01. The initial metallicity, $Z_{\rm init}$, that we input to MESA is consistent with the parameters used for isochrone fitting, as outlined in Table \ref{T3}. As described in Section \ref{S1}, the morphology of the HB is significantly influenced by the stellar mass loss rate, $\eta$. This mass loss rate, in turn, depends on the metallicity of the cluster and the generation of the stellar population to which the stars belong \citep{Tail20a}; thus, it remains highly uncertain in our simulations. The four star clusters we examined have metallicities ranging from $Z$=0.0002 to 0.0005, corresponding to [Fe/H]$\sim$$-$1.9 to $-$1.5. According to \cite{Tail20a}, first-generation stars within this metallicity range exhibit mass loss rates, $\eta$$\sim$0.25 to 0.35. Consequently, for our calculations, we simplistically adopted $\eta$=0.3, solely to qualitatively analyze the differences in He abundance distributions derived from the HB compared to those obtained from the MS. 

\begin{figure}[ht!]
\begin{center}
\includegraphics[width=0.5\textwidth]{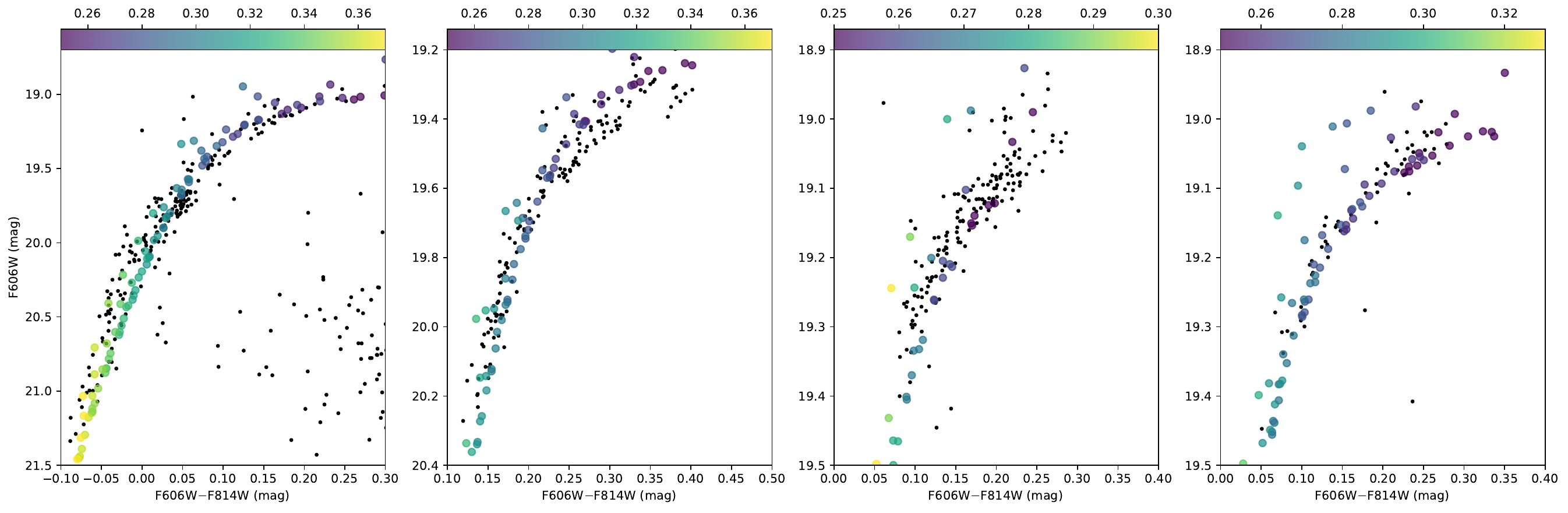}
  \caption{Comparison of observed and simulated HB. From left to right are the observed and simulated HBs for clusters Hodge 11, NGC 1841, NGC 2210 and NGC 2257. color bars indicate the helium mass fraction, $Y$, of the simulated HB stars.}
  \label{F9}
\end{center}
\end{figure}

Our results are shown in Figure \ref{F9}, and we find that the He abundance distributions obtained from the HB of the four star clusters do not always match those obtained from analyzing the MS. By visually comparing the observed HB morphology with the predictions from our models, we discovered that the two clusters, Hodge 11 and NGC 1841, may exhibit helium spreads of approximately $\delta{Y}\sim0.12$ and 0.09, respectively. This is consistent with the He distribution obtained from analysis of their MS morphology ($\delta{Y}\sim0.10-0.12$ and $\delta{Y}\sim0.08-0.11$, respectively). However, the distributions of $\delta{Y}$ obtained from fitting the HB morphology of the other two clusters, NGC 2210 and NGC 2257, are very different from the results obtained from fitting the MS. The fitting results of the HB indicate that the helium abundance variation in NGC 2210 and NGC 2257 is around $\delta{Y}\sim0.03-0.05$, which is only one-third to half of the results obtained from fitting the MS morphology. It is clear from Figure \ref{F9} that the HB of NGC 2210 and NGC 2257 is noticeably shorter than that of Hodge 11 and NGC 1841. Moreover, Figure \ref{F8} shows that the MS width of NGC 2210 and NGC 2257 is similar to Hodge 11 and is even a bit wider than that of NGC 1841's MS. 

\section{Discussions and conclusions}\label{S4}

We first discuss the reliability of our results. Many factors can contribute to the broadening of the MS. Besides the helium abundance dispersion, observational factors include photometric errors and PSF residuals. Physical factors include unresolved binaries, line-of-sight blending caused by crowding, differential extinction, dispersion of light elements (usually C, N, and O), metallicity dispersion, and field contamination. We discuss them separately. 

{\sc Dolphot} automatically provides magnitude uncertainties; however, these uncertainties are often inaccurate as they only account for the magnitude uncertainty arising from the counting uncertainty. For fainter sources, count uncertainty is the dominant source of error, while for the majority of brighter sources, it typically underestimates the true observational errors. Using ASs to simulate stellar populations can effectively avoid underestimating the errors mentioned above. This is because ASs are created based on the PSF fitted to observations, and their magnitudes are measured using photometry that matches the real observations. Therefore, in principle, ASs can account for errors from instrument effects such as bad pixels or saturation. Additionally, we add ASs in batches to the original frame for photometry, which allows us to consider errors from crowding, cosmic rays, and variation of background. In fact, by comparing the measured magnitudes of ASs to the input magnitudes, we find that the discrepancies are indeed greater than the theoretical errors provided by {\sc Dolphot}.

PSF residuals not only depend on the PSF used, but often differ across different passbands, leading to effects similar to differential extinction and causing color spread to be greater than theoretically expected. However, since the direction of broadening caused by PSF residuals is not perpendicular to that of differential reddening, correcting for differential reddening automatically accounts for zero-point variations caused by PSF residuals. Therefore, we did not further iterate in the color direction to correct for the PSF residuals. We compared the results of correcting only for differential reddening and only for PSF residuals, and found that the color dispersion we obtained was nearly identical in both amplitude and spatial distribution.

The observational effects caused by unresolved binaries and blending are essentially the same (if both sources of blending are stars). ASs can statistically simulate the blending effect, but the impact of unresolved binaries cannot be eliminated. However, since both blending and unresolved binaries only shift photometric results towards the red end of the MS, analyzing only the blue half of the MS can help minimize the influence of binaries and blending.

If MPs are present, different stellar populations may have different light elements. Apart from He, the greatest variations are often seen in C, N, O, and Na, and changes in these elements can significantly affect photometry. According to \cite{Cass17a}, C, N, O, and Na can produce significant bolometric corrections in the blue and UV passbands, while their impact in the optical bands is moderate, especially having an almost negligible impact on the F606W$-$F814W color band.

Through the fitting of isochrones, we can rule out the presence of metallicity spread, for reasons similar to those presented in \cite{Li23a}. Changes in [Fe/H] would not only affect the dwarf stars but would also impact the colors of giants. Nevertheless, we do not observe any significant broadening of the RGB. Furthermore, we confirm that variations in [$\alpha$/Fe] have nearly no impact on photometry.

We argue that the largest uncertainty in the broadening of the MS comes from field star contamination. Because the Magellanic Clouds are very far away, past attempts to decontaminate field stars have mostly used statistical methods. Specifically, this involves selecting a reference field in the outer areas of the FoV, far from the cluster center, or performing a parallel observation that is far away from the cluster involving the same passbands. It is assumed that the stars in these outer regions are mostly field stars. Then, a comparison of the CMDs between the outskirt and center regions of the cluster is made, followed by a statistical random removal of a certain number of stars from different parts of the diagrams \citep[e.g.,][]{Li16a}. Although this method can in principle statistically estimate the shape of the cluster's CMD, it cannot determine whether specific stars are members of the cluster, which often leads to controversy \citep[e.g.,][]{Cabr16a}. Using stellar PM to determine whether a star belongs to the cluster is the only way to avoid such controversy \citep[e.g.,][]{Wang24a}. 

We have confirmed that the use of previous statistical methods for field star decontamination does not significantly change the result that the MSs of the four clusters are significantly broadened \citep[See,][]{Li23a}. However, the seven-year observation interval may only allow for a limited distinction between cluster member stars and field stars. Therefore, we expect that the so-called `cluster sample' may still contain some field stars. Our fitting of the radial number density in the field of these clusters indicates that the contamination rate from field stars does not dominate. They only account for only 5\% to 25\% of the entire population. Therefore, even if field stars are not removed, we should expect at least three quarters of the dwarfs would distribute within a narrow MS, a typical SSP signal that cannot be obscured by a small number of field stars. However, the MS we observed is nearly uniformly distributed over a wide color range, suggesting that field star contamination alone cannot account for the observed MS morphology. 

\begin{figure}[ht!]
\begin{center}
\includegraphics[width=0.5\textwidth]{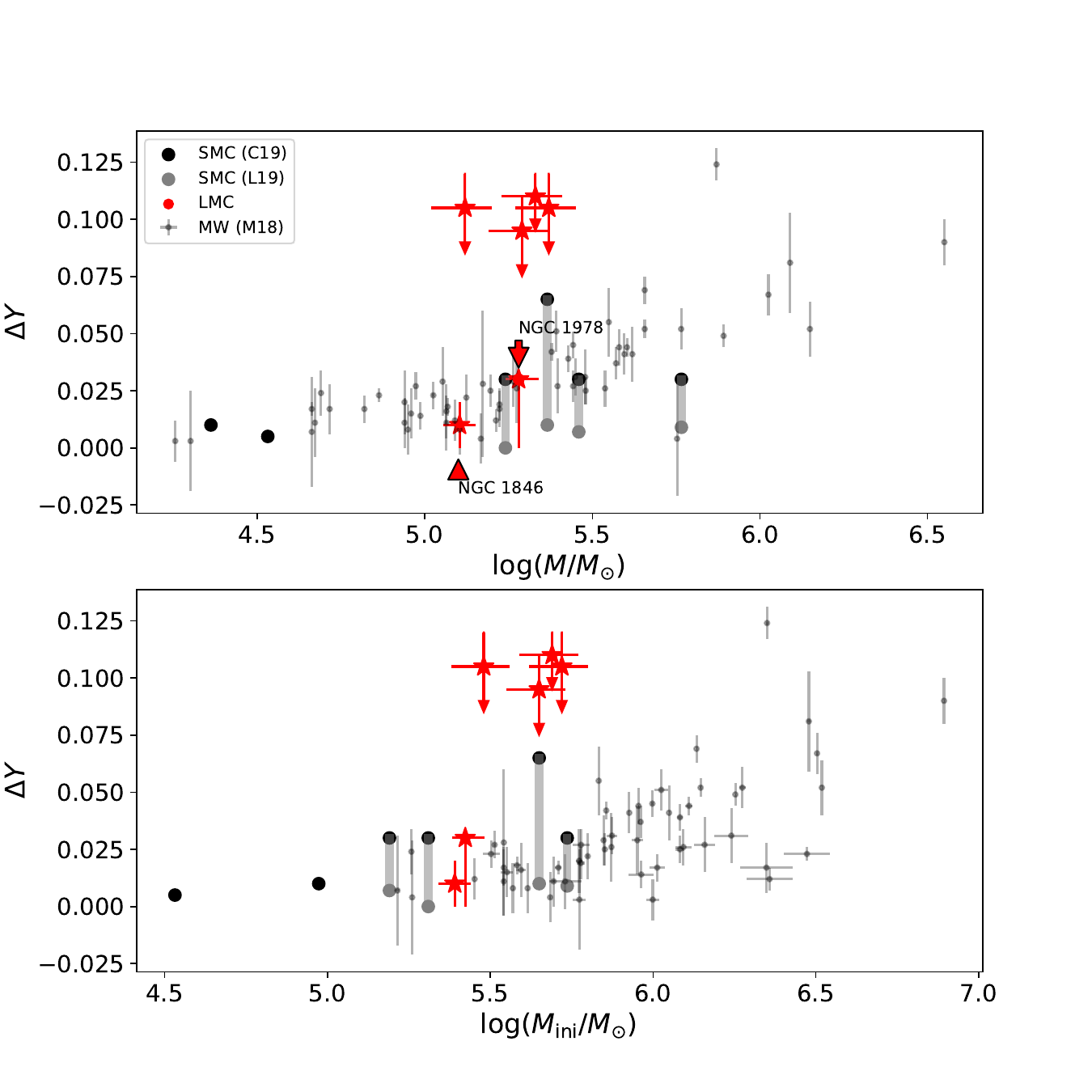}
  \caption{Internal helium spread and cluster mass relationships. Top: the internal helium spreads and the clusters present-day masses diagram, $\delta{Y}$--$\log(M/M_{\odot})$. Red  pentagrams are LMC clusters. Small grey dots are Milky Way GCs. Dark/light grey circles are SMC clusters. Bottom: the correlation between the internal helium spreads and the clusters initial masses, for Milky Way GCs, LMC and SMC clusters. The arrow indicates that we consider this value to represent only an upper limit. The arrow indicates that we consider this value to represent only an upper limit.}
  \label{F10}
\end{center}
\end{figure}

We acknowledge that some of our earlier conclusions about NGC 2210 are no longer reliable because we did not recognize the significant photometric errors in the central region of the cluster. Although we still stand by our main conclusion that NGC 2210 shows a significant broadening of the MS, our claim that its central region may possess a higher proportion of 2P stars is inaccurate \citep[][their fig.9]{Li23a}. Our conclusion was based on the observation of a greater number of MS stars with significant color dispersion in the center of the cluster, which we interpreted as indicating higher He abundance. However, this may simply result from larger photometric errors in the stars of the central region. We apologize for this error. Although in this work, we have excluded the very central region with larger photometric errors and only analyzed the relative outskirts of the cluster ($>10$ pc to the cluster center), where the photometric errors remain moderate and stable. However, this does not mean that crowding has no effect in this region; it only indicates that the contribution of crowding has become stable. The ${\delta{Y}}$ obtained by our method might only represent an upper limit. Similar to \cite{Li23a}, we present our results using Galactic GCs \citep{Milo17a}, two younger LMC clusters \citep{Li21a, Ji22a}, and SMC clusters \citep{Chan19a, Lagi19a} on the cluster mass -- $\delta{Y}$ diagram. From Figure \ref{F10}, it is evident that the four clusters we measured appear as outliers in the mass -- $\delta{Y}$ diagram, independent of whether we use their present masses (top panel) or initial masses (bottom panel, see \cite{Baum19a} for the calculation of initial masses). Next-generation instruments with higher spatial resolution may further refine our results.

The differences in results obtained from fitting the morphology of the HB and the MS clearly highlight the importance of mass loss. For example, the mass loss rate varies among different clusters and also between 1P and 2P population stars, even within the same cluster \citep{Tail20a}. If a star is in a close binary system, the companion's tidal forces will also affect its mass loss rate. A rapidly rotating star (for example, due to tidal locking that enhances the spin of a red giant) may have an extended red-giant lifetime, leading to greater mass loss, which results in its position in a bluer end of the HB. Overall, the complex evolutionary pathways of stars can result in different luminosities and colors during their HB phase, suggesting that the characteristics of the HB should not be the only criterion for evaluating a star's He abundance. We also emphasize that Hodge 11 may be a particularly interesting LMC GC similar to the Galactic GC, NGC 2808, as its extended HB and broadened MS are consistent well with models exhibiting significant He enrichment ($\delta{Y} > 0.10$).

If the estimated He abundance can represent the average level of the entire clusters, then the He enrichment in these four LMC clusters is comparable to that of the most massive GC in the Galaxy ($>10^6$ M$_{\odot}$). This is surprising, as the current masses of these clusters are between $1-2.5\times10^5$ M$_{\odot}$ \citep{Mcla05a}. The $f_{\rm 2P}$ for these four clusters are quite similar, therefore we did not observe a clear correlation between $f_{\rm 2P}$ and their present-day masses. This is similar to the results found by \cite{Milo20a} in intermediate-age clusters of the Magellanic Clouds (illustrated in their Figure 7, top left). However, this outcome might also stem from the small size of our cluster sample, which could introduce variability. The derived $f_{\rm 2P}$ for our four clusters ranges from $\sim$40\%$-$90\%, depending on the chosen He distribution models. {If present}, these values would be comparable to those observed in Galactic GCs and are higher than those observed in intermediate-age clusters in the Magellanic Clouds, as reported by \cite{Milo20a}. These four clusters are dominated by 2P stars, similar to the fact that extreme He-enriched clusters in the Milky Way are primarily occupied by 2P stars. {This also differs from the old GCs in the SMC, where first-population stars are more dominant in number \citep[e.g.,][]{Dale16a,Lagi19a}.} This may suggests that the MPs in these four LMC clusters may have origins similar to those of Milky Way GCs \cite{Milo18a}, {which is hard to explain because LMC clusters likely had weaker interactions with their host galaxy, keeping most of their original mass. This makes it unlikely for them to form such high fractions of second-population stars \citep{Parm24a}. Again, given the potential effects of crowding, it remains challenging to assess the extent to which this result requires further refinement, however.}

{Assuming that our results accurately represent the helium spread within the cluster, they may provide insights into various MPs models.} For the bimodal He distribution, our results indicate that the $\delta{Y}$ for the four clusters ranges from 0.08 to 0.10. Current mainstream models can generally produce such a significant He enrichment; however, VMS models tend to yield smaller helium enrichments (with $\delta{Y} < 0.01$). The observed $\delta{Y}$ may suggest that the wind material from VMS originates from their later evolutionary stages \citep[][their Section 4.3.2]{Giel18a}. However, if the actual He distribution is uniform, our results indicate that at least three clusters (Hodge 11, NGC 2210, NGC 2257) would have $\delta{Y}$ of at least 0.12. This approaches the maximum He enrichment limit that AGB models can produce \citep[0.36$-$0.38,][]{Sies10a,Dohe14a}. This suggests that AGB stars are unlikely to be the sole source of the observed He-enriched stars, or at least not the only contributors. FRMS models can generate significant He enrichment, reaching up to $Y$=0.8 \citep{Decr07a}, while massive IBs can, in principle, produce a wide range of He enrichments by adjusting binary parameters \citep{Demi09a}. If there are extreme He-enriched stars with $\delta{Y} > 0.12$, we may need to consider the contributions from FRMS or IBs.

In conclusion, we find that the LMC GCs Hodge 11, NGC 1841, NGC 2210, and NGC 2257 have MS that is significantly wider than expected from the SSP model, which cannot be explained by photometric uncertainties, field contaminations, unresolved binaries, differential reddening and metallicity spread. We suggest that they may all have significant He enrichment among their MS stars. If present, this enrichment is similar to what is observed in Galactic GCs that exhibit extreme variations in He abundance. However, owing to the great distance of the LMC, our results are likely to represent an upper limit.

\begin{acknowledgements}
This project is supported by the National Natural Science Foundation of China (NSFC grant Nos. 12033013 and 12073090), the National Key R\&D Program of China (2020YFC2201400) and the China Manned Space Project with NO.CMS-CSST-2021-A08, CMS-CSST-2021-B03.
\end{acknowledgements}

\bibliographystyle{aa}

\bibliographystyle{aasjournal}

\begin{thebibliography}{99}
\expandafter\ifx\csname natexlab\endcsname\relax\def\natexlab#1{#1}\fi
{\bibitem[Baumgardt et al.(2019)]{Baum19a} Baumgardt, H., Hilker, M., Sollima, A., et al.\ 2019, \mnras, 482, 5138. }

\bibitem[Bellini et al.(2013)]{Bell13a} Bellini, A., Piotto, G., Milone, A.~P., et al.\ 2013, \apj, 765, 32.

\bibitem[Cabrera-Ziri et al.(2016)]{Cabr16a} Cabrera-Ziri, I., Niederhofer, F., Bastian, N., et al.\ 2016, \mnras, 459, 4218. 

\bibitem[Cadelano et al.(2023)]{Cade23a} Cadelano, M., Pallanca, C., Dalessandro, E., et al.\ 2023, \aap, 679, L13.

\bibitem[Carini et al.(2020)]{Cari20a} Carini, R., Biazzo, K., Brocato, E., et al.\ 2020, \aj, 159, 152.

\bibitem[Cardelli et al.(1989)]{Card89} Cardelli, J.~A., Clayton, G.~C., \& Mathis, J.~S.\ 1989, \apj, 345, 245.

\bibitem[Carini et al.(2024)]{Cari24a} Carini, R., Sollima, A., Brocato, E., et al.\ 2024, \mnras, 528, 909.

\bibitem[Cassisi \& Salaris(2013)]{Cass13a} Cassisi, S. \& Salaris, M.\ 2013, Old Stellar Populations: How to Study the Fossil Record of Galaxy Formation, 538 pp.ISBN: 978-3-527-41059-0. Wiley-VCH, March 2013

\bibitem[Cassisi et al.(2017)]{Cass17a} Cassisi, S., Salaris, M., Pietrinferni, A., et al.\ 2017, \mnras, 464, 2341.

\bibitem[Chantereau et al.(2019)]{Chan19a} Chantereau, W., Salaris, M., Bastian, N., et al.\ 2019, \mnras, 484, 5236. 

\bibitem[Dalessandro et al.(2013)]{Dale13a} Dalessandro, E., Salaris, M., Ferraro, F.~R., et al.\ 2013, \mnras, 430, 459. 

{\bibitem[Dalessandro et al.(2016)]{Dale16a} Dalessandro, E., Lapenna, E., Mucciarelli, A., et al.\ 2016, \apj, 829, 77. }

\bibitem[Decressin et al.(2007)]{Decr07a} Decressin, T., Meynet, G., Charbonnel, C., et al.\ 2007, \aap, 464, 1029. 

\bibitem[de Mink et al.(2009)]{Demi09a} de Mink, S.~E., Pols, O.~R., Langer, N., et al.\ 2009, \aap, 507, L1.

\bibitem[D'Ercole et al.(2008)]{Derc08a} D'Ercole, A., Vesperini, E., D'Antona, F., et al.\ 2008, \mnras, 391, 825.

\bibitem[Doherty et al.(2014)]{Dohe14a} Doherty, C.~L., Gil-Pons, P., Lau, H.~H.~B., et al.\ 2014, \mnras, 441, 582.

\bibitem[Dolphin.(2011a)]{Dolp11a} Dolphin A., DOLPHOT/WFC3 user's
  guide, version 2.0. {\url
   {http://americano.dolphinsim.com/dolphin/dolphotWFC3.pdf}}

\bibitem[Dolphin.(2011b)]{Dolp11b} Dolphin A., DOLPHOT/WFPC2 user's
  guide, version 2.0. {\url
 {http://americano.dolphinsim.com/dolphot/dolphotWFPC2.pdf}}

\bibitem[Dolphin.(2013)]{Dolp13a} Dolphin A., DOLPHOT user's guide,
 version 2.0. {\url
   {http://americano.dolphinsim.com/dolphot/dolphot.pdf}}

\bibitem[Dupree \& Avrett(2013)]{Dupr13a} Dupree, A.~K. \& Avrett, E.~H.\ 2013, \apjl, 773, L28.

\bibitem[Gaia Collaboration et al.(2023)]{Gaia23a} Gaia Collaboration, Vallenari, A., Brown, A.~G.~A., et al.\ 2023, \aap, 674, A1.

{\bibitem[Parmentier(2024)]{Parm24a} Parmentier, G.\ 2024, \apj, 964, 140.}

\bibitem[Gieles et al.(2018)]{Giel18a} Gieles, M., Charbonnel, C., Krause, M.~G.~H., et al.\ 2018, \mnras, 478, 2461.

\bibitem[Gratton et al.(2004)]{Grat04a} Gratton, R., Sneden, C., \& Carretta, E.\ 2004, \araa, 42, 385. 

\bibitem[Gratton et al.(2013)]{Grat13a} Gratton, R.~G., Lucatello, S., Sollima, A., et al.\ 2013, \aap, 549, A41.

\bibitem[Grundahl et al.(1999)]{Grun99a} Grundahl, F., Catelan, M., Landsman, W.~B., et al.\ 1999, \apj, 524, 242. 

\bibitem[Hesser \& Bell(1980)]{Hess80a} Hesser, J.~E. \& Bell, R.~A.\ 1980, \apjl, 238, L149.

{\bibitem[Ji et al.(2022)]{Ji22a} Ji, X., Li, C.-Y., \& Deng, L.-C.\ 2022, RAA, 22, 035008. }

\bibitem[Jian et al.(2024)]{Jian24a} Jian, M., Fu, X., Matsunaga, N., et al.\ 2024, \aap, 687, A189.

\bibitem[Jiang et al.(2014)]{Jian14a} Jiang, D., Han, Z., \& Li, L.\ 2014, \apj, 789, 88. 

\bibitem[{{King}(1962)}]{King62a}
{King}, I. 1962, \aj, 67, 471,

\bibitem[Krause et al.(2013)]{Krau13a} Krause, M., Charbonnel, C., Decressin, T., et al.\ 2013, \aap, 552, A121. 

{\bibitem[Lagioia et al.(2018)]{Lagi18a} Lagioia, E.~P., Milone, A.~P., Marino, A.~F., et al.\ 2018, \mnras, 475, 4088.}

\bibitem[Lagioia et al.(2019)]{Lagi19a} Lagioia, E.~P., Milone, A.~P., Marino, A.~F., et al.\ 2019, \apj, 871, 140.

\bibitem[Lanzoni et al.(2019)]{Lanz19a} Lanzoni, B., Ferraro, F.~R., Dalessandro, E., et al.\ 2019, \apj, 887, 176.

\bibitem[Larsen et al.(2012)]{Lars12a} Larsen, S.~S., Strader, J., \& Brodie, J.~P.\ 2012, \aap, 544, L14. 

\bibitem[Li et al.(2020)]{Li20a} Li, L., Shao, Z., Li, Z.-Z., et al.\ 2020, \apj, 901, 49.

\bibitem[Li et al.(2021)]{Li21a} Li, Z.-Z., Li, L., \& Shao, Z.\ 2021, Astronomy and Computing, 36, 100483. 

\bibitem[Li et al.(2016)]{Li16a} Li, C., de Grijs, R., Deng, L., et al.\ 2016, \nat, 529, 502.

\bibitem[Li et al.(2023)]{Li23a} Li, C., Ji, X., Wang, L., et al.\ 2023, \apj, 948, 8.

\bibitem[Marino et al.(2014)]{Mari14a} Marino, A.~F., Milone, A.~P., Przybilla, N., et al.\ 2014, \mnras, 437, 1609.

\bibitem[McLaughlin \& van der Marel(2005)]{Mcla05a} McLaughlin, D.~E. \& van der Marel, R.~P.\ 2005, \apjs, 161, 304.

\bibitem[Milone et al.(2012)]{Milo12} Milone, A.~P., Piotto, G., Bedin, L.~R., et al.\ 2012, \aap, 540, A16. 

{\bibitem[Milone et al.(2017a)]{Milo17a} Milone, A.~P., Piotto, G., Renzini, A., et al.\ 2017, \mnras, 464, 3636. }

\bibitem[Milone et al.(2018)]{Milo18a} Milone, A.~P., Marino, A.~F., Renzini, A., et al.\ 2018, \mnras, 481, 5098.

\bibitem[Milone et al.(2020)]{Milo20a} Milone, A.~P., Marino, A.~F., Da Costa, G.~S., et al.\ 2020, \mnras, 491, 515. 

\bibitem[Milone \& Marino(2022)]{Milo22a} Milone, A.~P. \& Marino, A.~F.\ 2022, Universe, 8, 359.

\bibitem[Nardiello et al.(2015)]{Nard15a} Nardiello, D., Piotto, G., Milone, A.~P., et al.\ 2015, \mnras, 451, 312.

\bibitem[Nataf et al.(2011)]{Nata11a} Nataf, D.~M., Gould, A., Pinsonneault, M.~H., et al.\ 2011, \apj, 736, 94. 

\bibitem[Nguyen \& Sills(2024)]{Nguy24a} Nguyen, M. \& Sills, A.\ 2024, \apj, 969, 18.

\bibitem[Niederhofer et al.(2024)]{Nied24a} Niederhofer, F., Bellini, A., Kozhurina-Platais, V., et al.\ 2024, \aap, 689, A162.

\bibitem[Norris(2004)]{Norr04a} Norris, J.~E.\ 2004, \apjl, 612, L25.

\bibitem[O'Donnell(1994)]{Odon94} O'Donnell, J.~E.\ 1994, \apj, 422, 158.

\bibitem[Paxton et al.(2011)]{Paxt11a} Paxton, B., Bildsten, L., Dotter, A., et al.\ 2011, \apjs, 192, 3. 

\bibitem[Piotto et al.(2007)]{Piot07a} Piotto, G., Bedin, L.~R., Anderson, J., et al.\ 2007, \apjl, 661, L53. 

\bibitem[Siess(2010)]{Sies10a} Siess, L.\ 2010, \aap, 512, A10.

\bibitem[Smith(1987)]{Smit87a} Smith, G.~H.\ 1987, \pasp, 99, 67.

% \bibitem[Smith(2016)]{Smit16a} Smith, G.~H.\ 2016, \pasa, 33, e057.

\bibitem[Sbordone et al.(2011)]{Sbor11a} Sbordone, L., Salaris, M., Weiss, A., et al.\ 2011, \aap, 534, A9. 

\bibitem[Tailo et al.(2020)]{Tail20a} Tailo, M., Milone, A.~P., Lagioia, E.~P., et al.\ 2020, \mnras, 498, 5745. 

\bibitem[Valcarce et al.(2012)]{Valc12a} Valcarce, A.~A.~R., Catelan, M., \& Sweigart, A.~V.\ 2012, \aap, 547, A5.

\bibitem[Valcarce et al.(2013)]{Valc13a} Valcarce, A.~A.~R., Catelan, M., \& De Medeiros, J.~R.\ 2013, \aap, 553, A62. 

\bibitem[Vink(2018)]{Vink18a} Vink, J.~S.\ 2018, \aap, 615, A119.

\bibitem[Wagner-Kaiser et al.(2017)]{Wagn17a} Wagner-Kaiser, R., Mackey, D., Sarajedini, A., et al.\ 2017, \mnras, 471, 3347.

\bibitem[Wang et al.(2024)]{Wang24a} Wang, L., Deng, L., Pang, X., et al.\ 2024, \apj, 969, 21. 

\bibitem[Wang et al.(2020)]{Wang20a} Wang, L., Kroupa, P., Takahashi, K., et al.\ 2020, \mnras, 491, 440.

\bibitem[Winter \& Clarke(2023)]{Wint23a} Winter, A.~J. \& Clarke, C.~J.\ 2023, \mnras, 521, 1646. 

\end{thebibliography}

\begin{appendix}
\section{Datasets}
This section provides detailed information about the {HST} datasets used in this work. Table \ref{T2} includes the observational data we used, including the bands, exposure time, observation time, and Program ID.
\begin{table*}
\caption{Detail information of the {HST} datasets used in this work}
\label{T2}
\centering
\begin{tabular}{cccccc}
\hline\hline
Cluster & Instrument & Filter & Exposure Time & Start Time & Program ID \\
\hline
Hodge 11 & ACS/WFC & F606W & 2$\times$50 s + 6$\times$345 s + 6$\times$370 s = 4390 s & 2016/06 & 14164\\
  & ACS/WFC & F814W & 2$\times$70 s + 6$\times$345 s + 6$\times$377 s + 6$\times$410 s = 6932 s & 2016/07 & 14164\\
  & UVIS/WFC3 & F814W & 2$\times$40 s + 3$\times$446 s + 2$\times$447 s = 2312 s & 2022/03 & 16748\\\hdashline
NGC 1841 & ACS/WFC & F606W & 2$\times$50 s + 12$\times$353 s = 4336 s & 2015/12 & 14164\\
& ACS/WFC & F814W & 2$\times$70 s + 6$\times$352 s + 6$\times$385 s + 6$\times$420 s = 6962 s & 2015/12 & 14164\\
  & UVIS/WFC3 & F814W & 2$\times$40 s + 5$\times$456 s = 2360 s & 2022/01 & 16748\\\hdashline
NGC 2210 & ACS/WFC & F606W & 2$\times$50 s + 6$\times$348 s + 6$\times$353 s = 4306 s & 2016/11 & 14164\\
& ACS/WFC & F814W & 2$\times$70 s + 6$\times$344 s + 6$\times$378 s + 6$\times$413 s = 6950 s & 2016/11 & 14164\\
& UVIS/WFC3 & F814W & 2$\times$40 s + 3$\times$446 s + 2$\times$447 s = 2312 s & 2022/12 & 16748\\\hdashline
NGC 2257 & ACS/WFC & F606W & 2$\times$50 s + 6$\times$353 s + 3$\times$364 s + 2$\times$525 s = 4360 s & 2016/02 & 14164\\
 &ACS/WFC & F814W & 2$\times$70 s + 6$\times$363 s + 3$\times$390 s + 6$\times$400 s + 2$\times$450 s = 6788 s & 2016/02 & 14164\\
 & UVIS/WFC3 & F814W & 2$\times$35 s + 3$\times$433 s + 2$\times$434 s = 2237 s & 2022/03 & 16748\\
\hline
\end{tabular}
\end{table*}

\section{Field decontaminated color-magnitude diagrams}
This section presents the field decontaminated CMDs of three other GCs: NGC 1841, NGC 2210, and NGC 2257. The decontaminated CMD of Hodge 11 is provided as an example in Figure \ref{F2}, and relevant descriptions of these CMDs can be found in Section \ref{PM}. 
\begin{figure}[ht!]
\begin{center}
\includegraphics[width=0.5\textwidth]{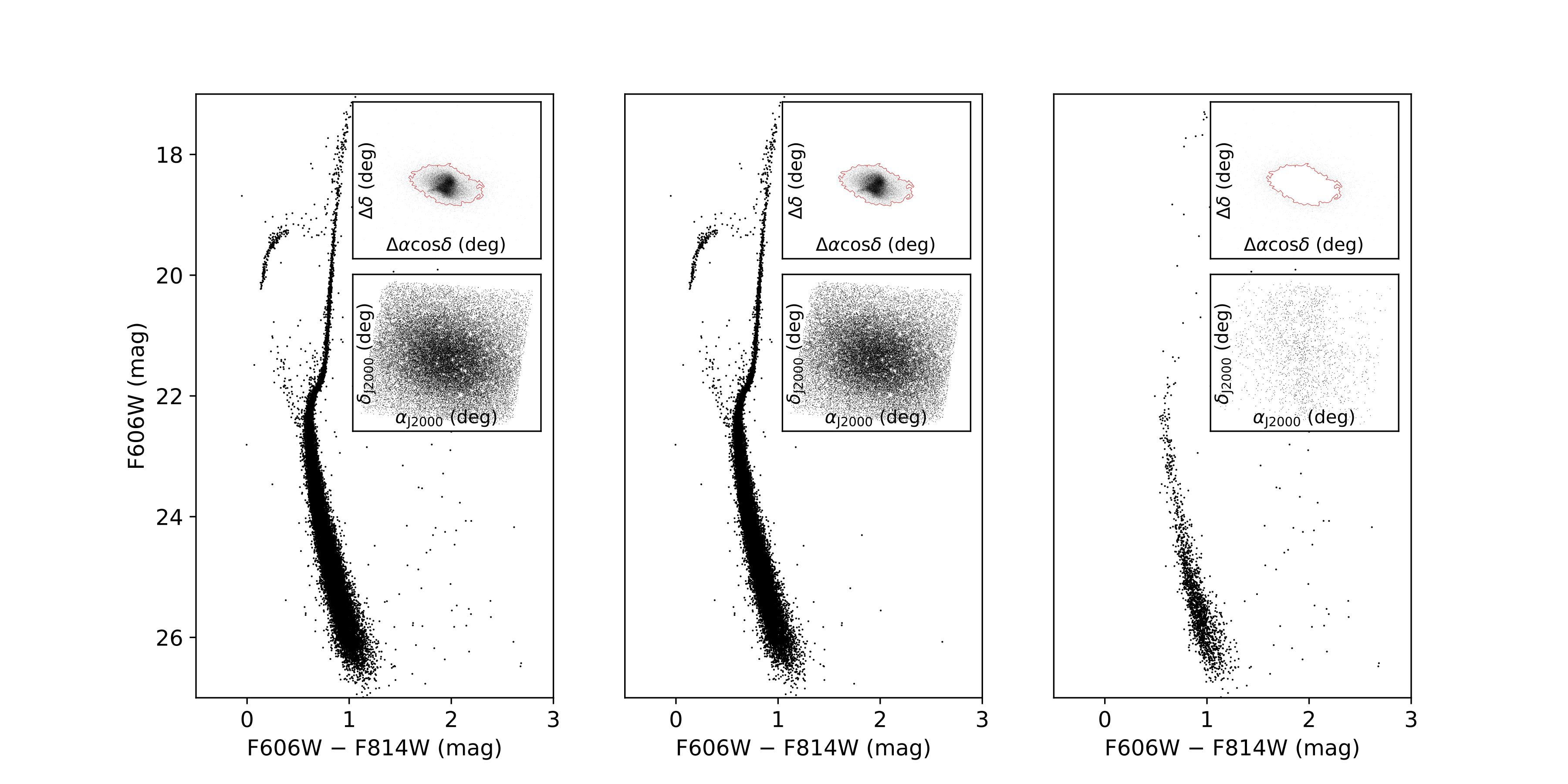}
  \caption{Same as in Fig. \ref{F2}, but for the cluster NGC 1841.}
  \label{F3}
\end{center}
\end{figure}  

\begin{figure}[ht!]
\begin{center}
\includegraphics[width=0.5\textwidth]{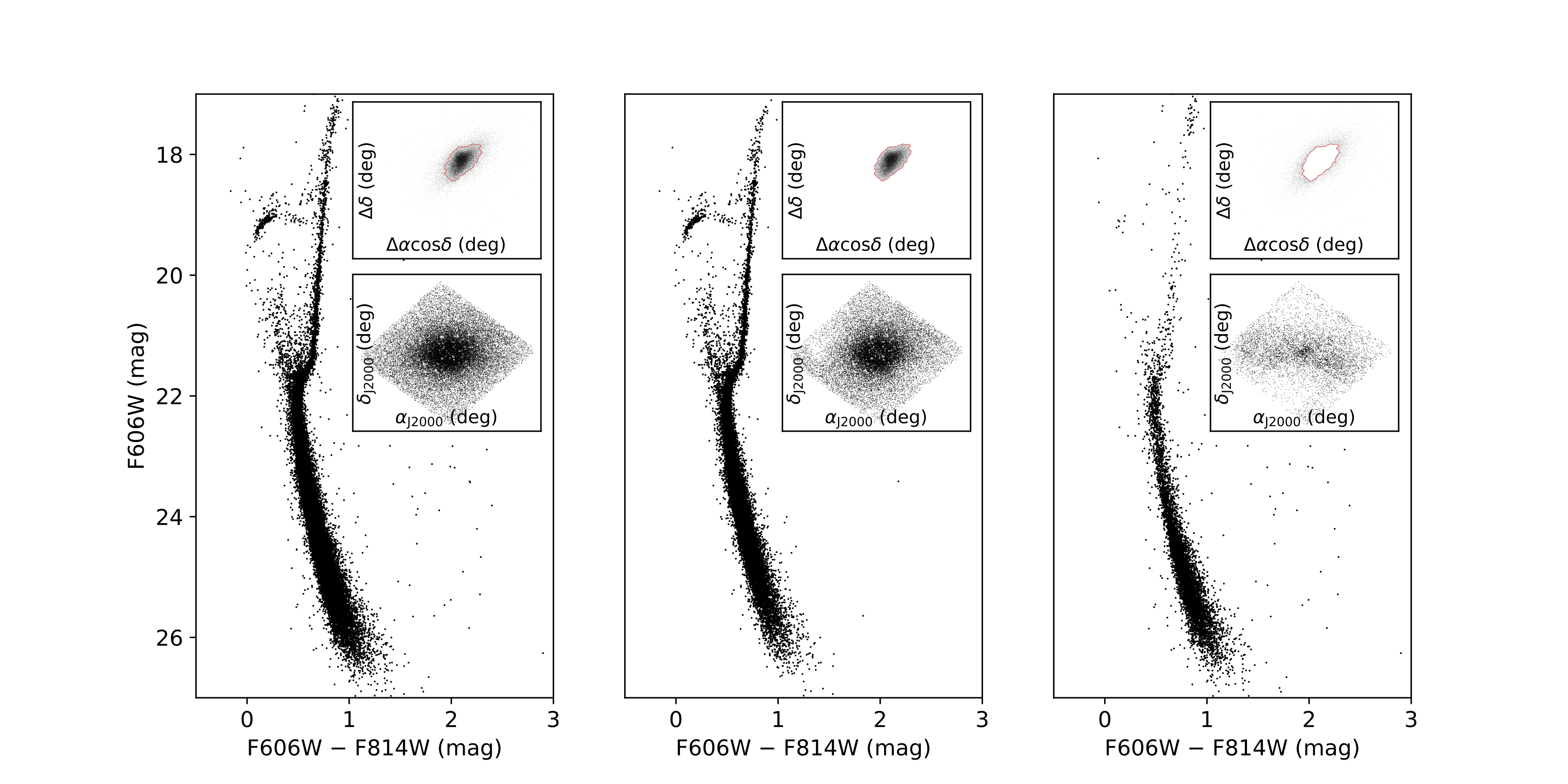}
  \caption{Same as in Fig. \ref{F2}, but for the cluster NGC 2210.}
  \label{F4}
\end{center}
\end{figure}  

\begin{figure}[ht!]
\begin{center}
\includegraphics[width=0.5\textwidth]{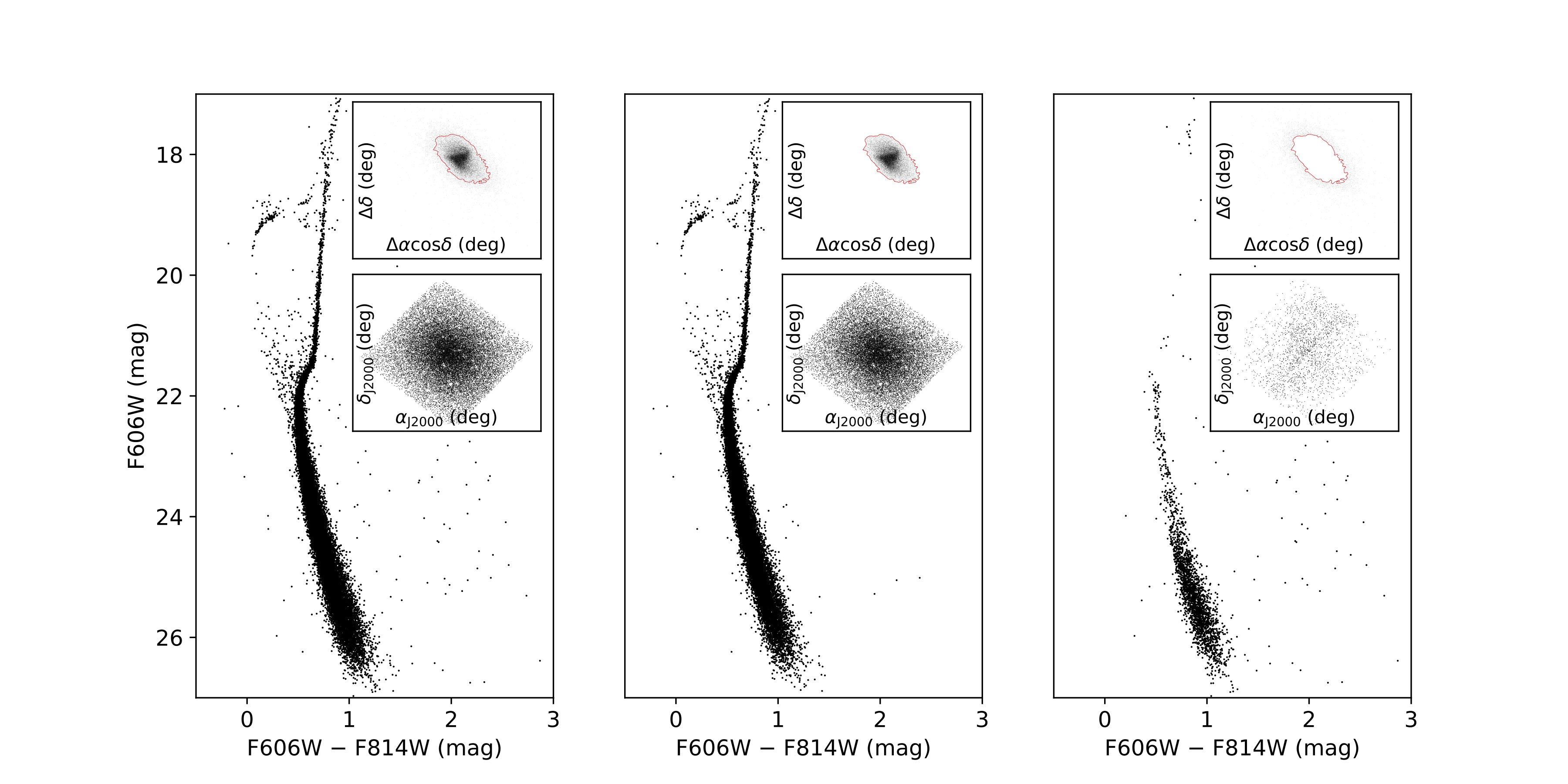}
  \caption{Same as in Fig. \ref{F2}, but for the cluster NGC 2257.}
  \label{F5}
\end{center}
\end{figure}   

\section{Photometric uncertainties as a function of radius}
\begin{figure}[ht!]
\begin{center}
\includegraphics[width=0.5\textwidth]{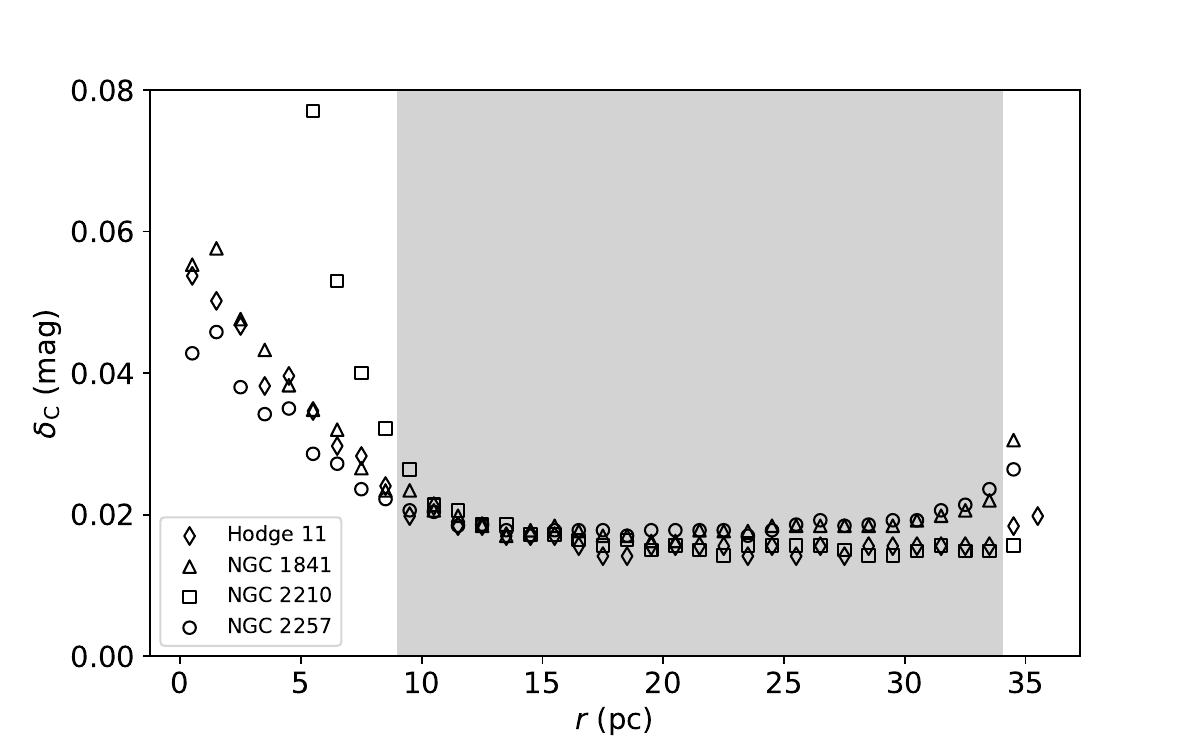}
  \caption{Color uncertainties as a function of radius from the cluster center
This figure depicts color uncertainties plotted against the radius from the center of the cluster. The shaded region highlights the radius range where color uncertainties remain approximately constant.}
  \label{A1}
\end{center}
\end{figure}

As shown in Figure \ref{A1}, due to crowding, the photometric errors increase sharply as the radius decreases within about 38 arcsec from the center of the cluster. Because of this, we select only stars within the range of 10 to 30 pc for the analysis in this work. We evaluate the photometric uncertainties near the MS by calculating the absolute value of the difference between the measured magnitudes of detected ASs and their input magnitudes. We calculate the corresponding color errors (as shown on the vertical axis of Figure A1) by adding the squared errors of the two passbands and then taking the square root. 

\section{Extinction residuals}
\begin{figure}[ht!]
\begin{center}
\includegraphics[width=0.5\textwidth]{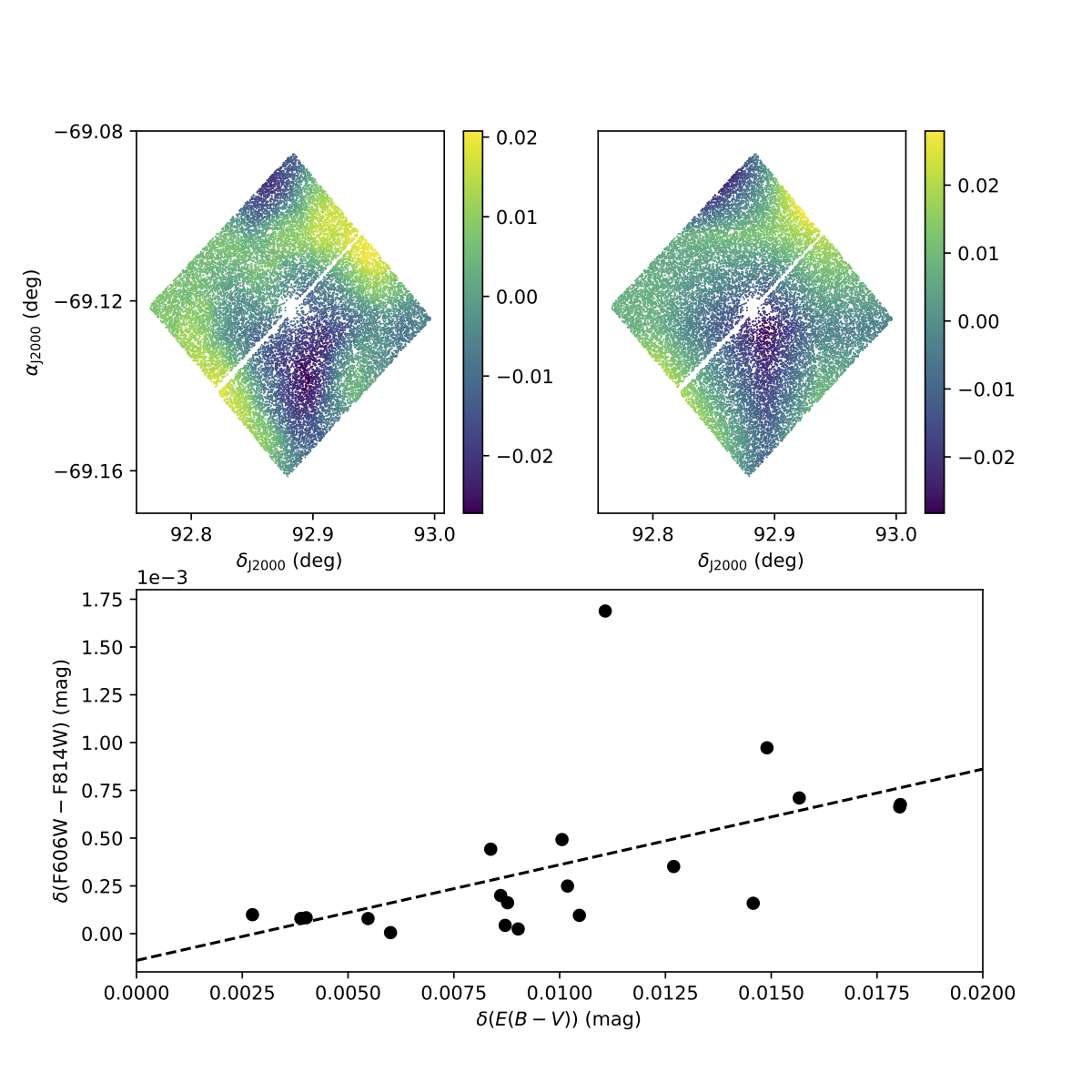}
  \caption{Deriving differential reddening residuals. The upper panels show the distribution of simulated stars with differential extinctions (upper-left), and with their differential extinction distribution derived from statistical methods (upper-right, see Section 2.5). The lower panel displays the correlation between the color residuals (after we corrected the differential extinction) and the values of differential extinctions.}
  \label{A2}
\end{center}
\end{figure}

Figure \ref{A2} shows an example we used to evaluate the differential extinction. The top left panel shows an input differential extinction spatial distribution map based on the ASs (with colors representing their extinction values), while the top right panel shows the derived differential extinction map using the statistical methods described in Section \ref{reddening}. The difference between these two maps gives us the extinction residuals left by our method. Based on this approach, we evaluate the most likely color residuals when obtaining different differential extinctions, and these results are subsequently used to correct our ASs photometry.

\label{lastpage}
\end{appendix}
\end{document}